\documentclass{article}
\usepackage[utf8]{inputenc}
\usepackage[margin=1in]{geometry}
\usepackage{subcaption}
\usepackage{graphicx}
\usepackage{svg}
\usepackage{amsmath}
\usepackage{amssymb}
\usepackage[hyphens]{url}
\usepackage{hyperref}
\hypersetup{colorlinks=false,breaklinks=true,hidelinks}
\usepackage{cleveref}
\usepackage{apacite}
\usepackage{natbib}
\usepackage{authblk}
\usepackage{multirow}
\usepackage{booktabs}
\usepackage{makecell}
\graphicspath{{figures/}}
\usepackage{setspace}
\title{QuakeFormer: A Uniform Approach to Earthquake Ground Motion Prediction Using Masked Transformers}
\author[1,2]{Yitian Feng}
\author[1,3]{Weiqiang Zhu\thanks{zhuwq@berkeley.edu}}
\author[2]{Xinzheng Lu\thanks{luxz@tsinghua.edu.cn}}
\affil[1]{Department of Earth and Planetary Science, University of California, Berkeley, CA, USA
}
\affil[2]{Department of Civil Engineering, Tsinghua University, Beijing, China}
\affil[3]{Berkeley Seismological Laboratory, University of California, Berkeley, Berkeley, CA, USA}
\date{}

\begin{document}

\maketitle

\begin{abstract}
Ground motion prediction (GMP) models are critical for hazard reduction before, during and after destructive earthquakes. 
In these three stages, intensity forecasting, early warning and interpolation models are corresponding employed to assess the risk.
Considering the high cost in numerical methods and the oversimplification in statistical methods, deep-learning-based approaches aim to provide accurate and near-real-time ground motion prediction.
Current approaches are limited by specialized architectures, overlooking the interconnection among these three tasks. 
What's more, the inadequate modeling of absolute and relative spatial dependencies mischaracterizes epistemic uncertainty into aleatory variability. 
Here we introduce QuakeFormer, a unified deep learning architecture that combines these three tasks in one framework. 
We design a multi-station-based Transformer architecture and a flexible masking strategy for training QuakeFormer. 
This data-driven approach enables the model to learn spatial ground motion dependencies directly from real seismic recordings, incorporating location embeddings that include both absolute and relative spatial coordinates.
The results indicate that our model outperforms state-of-the-art ground motion prediction models across all three tasks in our research areas.
We also find that pretraining a uniform forecasting and interpolation model enhances the performance on early warning task.
QuakeFormer offers a flexible approach to directly learning and modeling ground motion, providing valuable insights and applications for both earthquake science and engineering.

\end{abstract}

 \section{Introduction}
Ground motion is a fundamental aspect that bridges the fields of earthquake science and engineering. Ground Motion Prediction (GMP) models, which provide ground motion's intensity measures (IMs) for specific locations, play a vital role in various applications.
Before an earthquake occurs, ground motion forecasting models \citep{douglas2003earthquake}, including ground motion prediction equations(GMPEs) \citep{Abrahamson2014ASK14}, are essential for estimating its potential impact, allowing for seismic hazard analysis \citep{baker2013introduction}. 
During an impending earthquake, Earthquake Early Warning (EEW) systems  trigger alerts based on fast P-wave observations, thereby reduce potential risks and improve damage assessment \citep{cremen2020earthquake}. 
And after earthquake disasters, interpolation approaches, such as ShakeMap \citep{wald2022shakemap}, are employed to assess ground shaking at target sites using available observations, which helps emergency rescue and disaster recovery. 
The framework of existing EEW systems could be further categorized into source-based and propagation-based. The former  firstly estimates source properties, then derives ground shaking from forecasting models (usually GMPEs) \citep{chung2019optimizing,kohler2020earthquake,bose2012real,bose2018finder}; The latter uses interpolation models to predict shaking from the first image of the seismic wavefield \citep{kodera2018propagation}.

While physical-driven simulation models can deliver highly accurate GMP under specific scenarios \citep{graves2011cybershake}, their complexity often requires significant computational resources, which may not align with the urgent demands of rapid decision-making.
As an alternative, empirical formulas can quickly estimate location \citep{nakamura2007uredas, melgar2012real}, magnitude \citep{kuyuk2013global,tsang2007magnitude} and shaking \citep{Abrahamson2014ASK14}, making them widely utilized in early warning systems and seismic hazard analyses. 
However, the use of low-dimensional empirical formulas presents challenges in addressing both temporally and spatially varying features, as well as unprecedented seismic observations.
For example, in 2023, over 700 k seismic records from 20 k seismic events were collected  from the Northern and Southern California Earthquake Data Center Catalog\citep{NCEDC2014dataset, center2013southern}. 
This vast dataset has prompted the adoption of deep learning techniques, which can automatically extract high-dimensional spatial and temporal features from large volumes of data \citep{reichstein2019deep}.
Many recent studies have investigated deep learning approaches for ground motion forecasting \citep{monterrubio2024machine, joshi2024application},  interpolation \citep{otake2020deep,jozinovic2020rapid} as well as EEW \citep{munchmeyer2021transformer, fayaz2023deep}. 

Despite noteworthy advancements, existing methodologies continue to encounter two primary challenges. 
First, the existing methods primarily focus on  specialized GMP tasks, which not only limit their universality, but also overlook the interconnection between these three tasks. 
Ground motion forecasting model, such as GMPEs, are used to provide the initial and final intensity predictions for ShakeMap and traditional source-based EEW systems \citep{wald2022shakemap,chung2019optimizing,kohler2020earthquake,bose2012real,bose2018finder}.
These multi-stage methods may suffer from cumulative error propagation, ultimately leading to inaccurate estimations.
Existing deep learning based EEW methods \citep{munchmeyer2021transformer,datta2022deepshake} enable to output end-to-end prediction of IMs. 
But due to the real-time dataset for training, they do not explicitly consider the forecasting models, which provide valuable source/path/site effects. 
Second, significant uncertainty in data-driven approaches arises from improper modeling of relative and  absolute spatial dependency. To date, no existing deep learning method have addressed these two critical aspects.
Some existing studies oversimplify relative spatial dependencies using Gaussian and stationary assumption, or compute spatial correlations solely based on the distance between sites \citep{lu2021regional,wald2022shakemap,mori2022ground,fayazrewfers,zhang2022spatiotemporal}, struggling to account for the complex spatial variations in real-world scenarios \citep{goda2008spatial}.
Additionally, many methods overlook the impact of absolute regional variance.
For example, traditional ergodic GMPEs , assumed to be applicable to any location within the broad tectonic category, enables to provide station global average of ground motion. 
But the increasing ground motion records indicate that a large aleatory variability exists between an observation and the global average.
Significant systematic effects associated with site and source locations \citep{sahakian2018decomposing} are incorrectly characterized into randomness, leading to overestimating IMs at low probabilities of exceedance in the seismic hazard analysis.
In contrast, non-ergodic models, in which these location-specific effects are modeled explicitly, could help reduce the aleatory variability.
To incorporate absolute location, many efforts have been made to regionalize areas by subdividing broad regions into a grids \citep{landwehr2016nonergodic, abrahamson2019probabilistic}, or by training models using location coordinates, such as latitude and longitude \citep{monterrubio2024machine}. 
However, the grid-based approach presents challenges in effectively capturing multi-scale spatial distributions, while training with low-dimensional coordinates has shown poor performance in addressing high-frequency spatial variations \citep{Mildenhall2020NeRF}.

In this work, we propose a uniform GMP architecture based on masked Transformer, $\textbf{QuakeFormer}$,  that outperforms specialized architectures across ground motion forecasting, interpolation and early warning tasks. 
The key components in QuakeFormer is a masked pre-training strategy \citep{devlin2018bert, he2022masked} that randomly masks some of input tokens (unit of information) during training, and the objective is to predict the original intensity distribution based on its task-specific observations.
We jointly pre-train a uniform ground motion model for forecasting and interpolation tasks using the entire dataset in California, after which we fine-tune it to the EEW task.
For spatial dependencies, we employ location embeddings for both absolute and relative spatial coordinates. 
First, coordinates are mapped into high dimension space before putting into the neural network, thereby providing a learning-friendly representation for downstream models \citep{Mildenhall2020NeRF,Mai2021ARO}. 
Second, we explicitly incorporate relative position into self-attention layers in the Transformer using Rotary Position Embedding \citep{su2024roformer}, which offers valuable supervision for dependency modeling while maintaining low computational cost.

We conduct comprehensive evaluation that test on unseen events as well as unseen sites to demonstrates QuakeFormer's applicability for the future event and for new regional estimation respectively. 
Four state-of-art specialized  GMP methods are compared with QuakeFormer: empirical GMPE for forecasting, MVN for interpolation, EPS (including ElarmS) for source-based EEW and PLUM for propagation-based EEW. 
The results indicate that QuakeFormer achieves the best overall performance.
Additionally, we compare two training strategies for the EEW task: training QuakeFormer for a single EEW model (\emph{EEW-single}) and finetuning the uniform model for EEW purpose (\emph{EEW-finetune}). The results demonstrate that pre-training a ground motion model significantly enhances performance for the EEW task.

\section{Methods}
\subsection{Ground Motion Prediction Tasks}
\label{Ground motion model}
In this research, QuakeFormer is designed to predict the distribution of peak ground motion acceleration $\mathbf{PGA}_{ij}$ for event $i$ and station $j$. In the context of EEW, source parameters will also be estimated.  The input and output for the three ground motion tasks can be found in \Cref{tab:input output}. 

The three ground motion tasks differ based on the available observations used to interpret the scenario-based source mechanics. We denote $E_i$ as the event descriptions, $O_i^c$ as complete site observations, including $\mathbf{PGA}$ and waveforms in the duration time, $O_i^r$ as real-time site observations, $S_j$ as site descriptions and $t_i$ as the P-wave arrival time for site $i$.
For the ground motion forecasting, $\ln{\mathbf{PGA}}_{ij}=f(E_i, S_j)$. In our research, $E_i$ consists of a simple description of earthquake event, such as location, depth and $\textbf{M}$, corresponding to point source estimation. 
For the interpolation, $\ln{\mathbf{PGA}}_{ij}=f(E_i,O_i^c, S_j)$. The inclusion of wave observations allows the model to capture a more comprehensive understanding of the fault characteristics. Given the availability of complete waveforms after an earthquake, the time window in $O_i^c$ is defined as  $[t_i-st, t_i+T]$, where $st$ is the analysis time before P-wave arrival , $T$ is a fixed duration time. 
And for the EEW task, $\ln{\mathbf{PGA}}_{ij}=f(E_i, O_i^r,S_j)$. The waveforms in $O_i^r$ are aligned by $t_0$,  the time of P-wave arrival at the first triggered station. The time window for EEW is defined as $[t_0-st, t_0+t]$, where $t$ is the time following the P-wave arrival, it also referred to as observation time. As earthquake evolves, the observation time $t$ increases.
Furthermore, EEW systems could be generally classified into on-site EEW and regional EEW. 
The former consists a limited seismic stations at or near the target sites, and predicts ground shaking within the system array. 
The latter consists of a network of seismic sensors and  predicts ground shaking in this region, including locations without seismic stations.

\begin{table}
\caption{Model input and output}
\label{tab:input output}
\centering
\begin{tabular}{@{}lcccc|cc@{}}
\hline
\toprule
              & \multicolumn{4}{c|}{Input}                              & \multicolumn{2}{c}{Output} \\
Task          & Event Description & Station Coordinate & Waveform & IMs & Event Description   & IMs\\ \midrule
Forecasting   & \checkmark                 & \checkmark                  &          &     &                     & \checkmark    \\
Interpolation & or                & \checkmark                  & complete& \checkmark   &                     & \checkmark    \\
Early Warning &                   & \checkmark                  & partial&     & \checkmark                   & \checkmark    \\ \bottomrule
\end{tabular}
\end{table}


\subsection{Dataset}
\begin{figure}[h]
    \centering
    \begin{subfigure}[b]{0.49\textwidth}
        \centering
        \includegraphics[width=\textwidth]{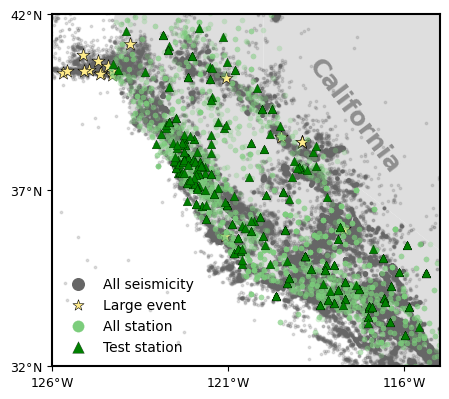}
        \caption{}
        \label{seismicity}
        
    \end{subfigure}
    \hfill
    \centering
    \begin{subfigure}[b]{0.49\textwidth}
        \centering
        \includegraphics[width=\textwidth]{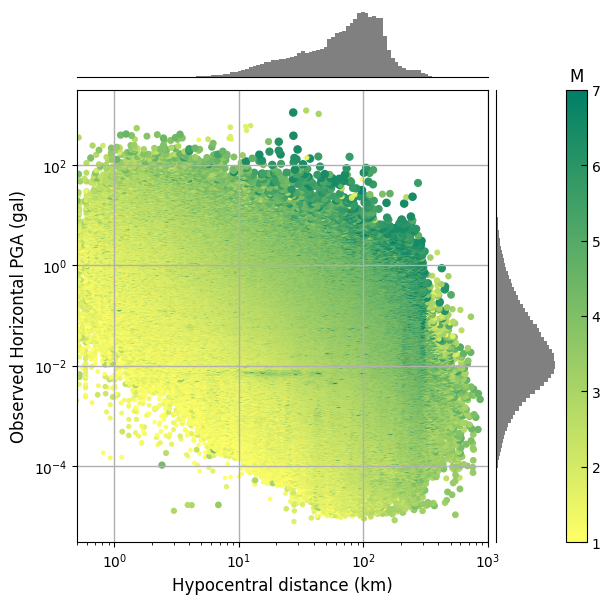}
        \caption{}
        \label{Mag-R}
    \end{subfigure}
    \caption{ (a) Maps of stations and earthquakes in the dataset. Light gray dots represent all seismicity we used, and yellow stars show the earthquake greater than M5.5. Green circles shows the seismic stations in the dataset, and the green triangles are stations belong to SR set. (b)Observed horizontal $\ln{PGA}$ in unit gal versus hypocentral distance $R_p$ in this dataset. The histograms above and on the left side are of $R_p$ and $\ln{PGA}$, respectively.}
    \label{Dataset}
    \
\end{figure}

In this paper, we utilize waveform data from Northern and Southern California Earthquake Data Center (NCEDC and SCEDC) \citep{NCEDC2014dataset, center2013southern}. After performing baseline correction, all waveforms are filtered by a 0.5 Hz high-pass Butterworth filter. The horizontal PGA, measured in units of gal, is calculated as the geometric mean of the horizontal components over the specified duration. 
We select dataset using the following criteria: 1) Earthquakes with $\textbf{M}<1$ are excluded; 2) Recordings of low-quality, such as signal noise ratio (SNR) of less tan 1 in any direction, as well as those affected by abnormal interference or exhibiting multiple jumps, are removed; 3) Recordings with exclusively low intensity, specifically those with $\mathbf{PGA}<10^{-6}$, are excluded; 4) Only surface seismograms are retained; 5) Earthquake with fewer than 5 recordings are excluded. 
After selection, our dataset contains over 4 million 3-component records belonging to 200,000 events in California from 2000 to 2024 (Fig.\ref{Dataset}). 

\subsection{Deep Learning Model}
\begin{figure}[h] 
    \centering
    \includegraphics[width=1\linewidth]{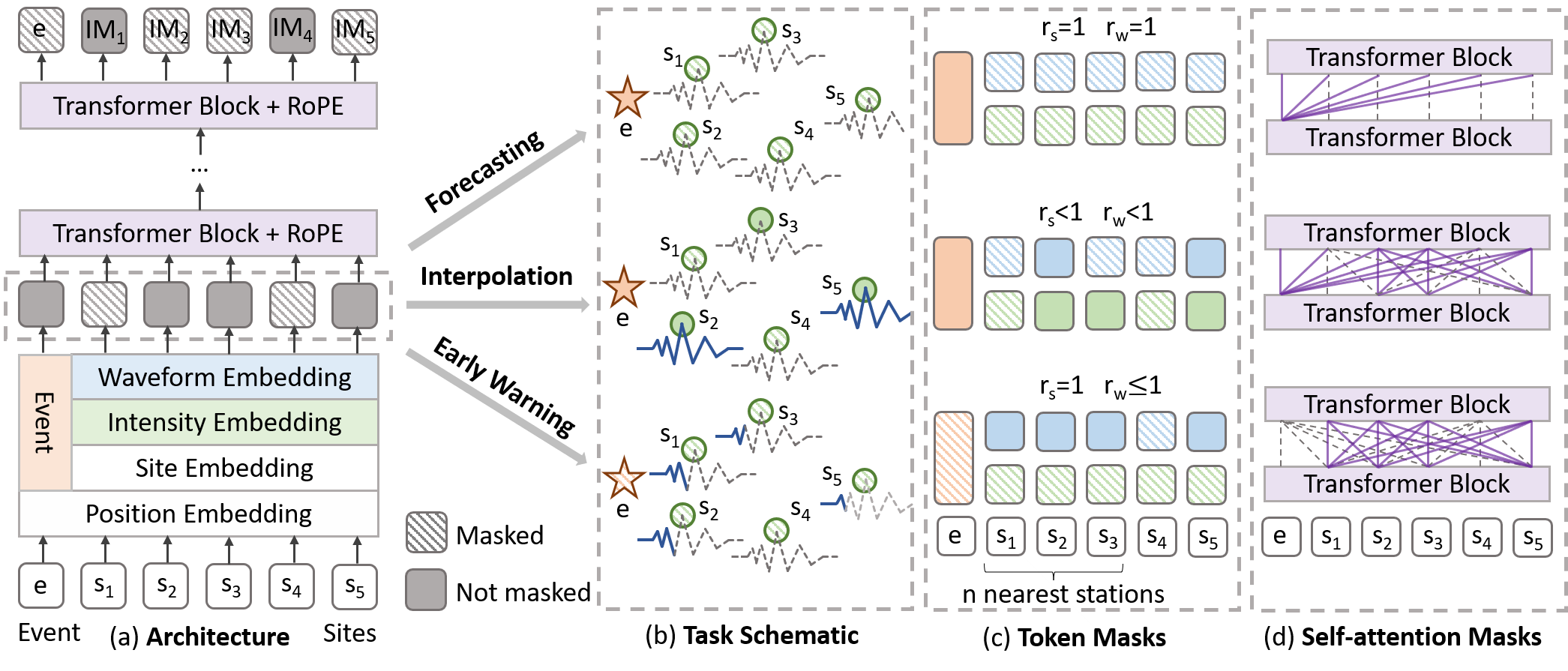}
    \caption{QuakeFormer Framework: (a) is the architecture of our model. The base model employs a multi-scale sinusoidal position encoder and a sequence encoders to obtain absolute position embeddings, site attribute embeddings, intensity embeddings and waveform embeddings. Mask strategy is then applied to randomly mask a portion of the intensity embeddings and waveform embeddings with ratio $r_s$, $r_w$, respectively. A sequence of Transfomer blocks with RoPE is utilized to predict the masked intensities and event token (in EEW). (b) is the schematic view of the station mask $M_S$, wave-station mask $M_{WS}$ and wave-time mask $M_{WT}$ for the three tasks. The solid-filled circles represent observed intensity values, while the circles filled with diagonal lines represent unobserved intensities at target sites (masked by $M_S$). The blue solid lines represent observed seismic waveforms, whereas locations denoted by dashed gray lines indicate unobserved waveforms(masked by $M_{WS}$ or partial-time waveforms(masked by $M_{WT}$. (c) shows the token masks in the architecture, corresponding to the three tasks. (d) presents the different self-attention masks that control the access to the updated tokens for each input tokens. Purple solid lines indicate tokens that are permitted to aggregate their representations with those of other tokens . Gray dashed lines represent tokens that can only accept the representation from other tokens, thereby establishing controlled interactions within the model.  }
    \label{fig:model_arch}
\end{figure}

We propose QuakeFormer, a straightforward encoder-only approach that reconstructs the shaking distribution based on task-specific observations. In this section, we introduce the base model, its important components (Section \ref{section:location embedding}), as well as its pre-train strategy (\Cref{section:masking strategy}). 
 \Cref{fig:model_arch} illustrates the model architecture. 

\subsubsection{Base model}
The input tokens in QuakeFormer represent a set of Points of  Interest $\mathcal{P}=\{p_i\}$, consisting of one event point and a sequence of site points. Each site point $p^s_i=(\mathbf{x}_i, \mathbf{v}_i, \mathbf{o}_i)$ is associated with a location $\mathbf{x}_i$ in 2-D space, attributes $\mathbf{v}_i$ (including $V_{S30}$, type of seismic instrument and elevation), and observations $\mathbf{o}_i$ ( including intensity $\mathbf{PGA}$ and waveforms). For simplicity, we incorporate the site elevation and epicenter depth into the attributes. Event point $p^e=(\mathbf{x}_i, \mathbf{v}_i)$ is invisible only for early warning task.
The main architecture consists three parts: feature extraction, observation masking and prediction.

In feature extraction, we encode location $\mathbf{x}_i$, attribute $\mathbf{v}_i$, intensity and waveform separately into latent embeddings of size $d$. Specifically, we use a multi-scale location encoder $Enc(\cdot)$ for locations (see Section \ref{section:location embedding}), use Conventional Neural Network (CNN) for 3-component waveforms, and apply Fully Connected Network (FCN) for attributes and intensity. For real-time applications in EEW, we randomly mask waveforms sequences in time before encoding them into wave embedding (Figure \ref{fig:model_arch} (a), Section \ref{section:masking strategy}).  More details about feature extraction could be seen in Appendix \ref{appendix:feature encoder}.

Following this, a masking strategy is applied to combine these three tasks in one architecture (see Section \ref{section:masking strategy}). In training, we randomly mask a varying fraction of latent intensity embeddings and waveform embeddings for each site, and replace them with special mask tokens, $\text{[IM]}$ and $\text{[WAVE]}$, to make them invisible. 
After masking, all embeddings for each point are summed up as input tokens, facilitating the fusion of modalities in the latent space. Similar to the $\text{[CLASS]}$ token used in BERT \citep{devlin2018bert}, we replace the input event token in EEW task with a special learnable embedding $\text{[EVENT]}$. Different site and event observations correspond to different tasks (Figure \ref{fig:model_arch} (b)). 

Then following the previous Transformer architecture\citep{vaswani2017attention}, we apply several Transformer encode layers, including self-attention blocks and MLP blocks, to extract features for the input tokens.
Rotary Position Embedding approach is applied in self-attention calculations to explicitly incorporate relative position dependency (see Section \ref{section:location embedding}).
Masked intensity embeddings and event description are predicted by the Transformer encoder and a FCN output layer.
To accommodate variable lengths of site token sequences for different events, we employ zero-padding to standardize these inputs to the same length. In the attention matrix, we set the values corresponding to the padding sections to zero to ensure accurate inference results. 
The loss is computed exclusively on the masked tokens  $\text{[IM]}$ and $\text{[EVENT]}$ (for EEW).

\subsubsection{Location Embedding}
\label{section:location embedding}

Despite neural networks are universal function approximators, operating the 2-D locations directly results in poor performance in representing high-frequency variability in geometry \citep{Mildenhall2020NeRF}. The probability of observations is characterized by a highly non-linear distribution. Projecting locations into a high-dimensional space before neural network could be interpreted as a feature decomposition process that makes learning friendly \citep{Mai2021ARO}. Besides, the physical process underlying ground motion involve a spatial distribution with very different scaling patterns. The finer spatial resolution leads to data sparsity and higher computation costs, while the sparser spatial resolution may yield poorer estimates. Therefore, we define an multi-scale location encoder
$Enc(\mathbf{x})=\mathbf{NN}(PE(\mathbf{x})):\mathbb{R}^2 \to \mathbb{R}^d  (2\ll d)$. Here, $\mathbf{NN(\cdot)}$ is a learnable shallow neural network, and $PE(\cdot)$ is a deterministic function that transforms location $\mathbf{x}$ into higher dimensional space $\mathbb{R}^{4N}$. N is the number of scale levels. Inspired by NeRF \citep{Mildenhall2020NeRF}, we use a sinusoidal multi-scale $PE(\cdot)$: 
\begin{equation}
    \label{location embedding}
    PE(\mathbf{x})=(\cos{2^{L_1}\mathbf{x}},\sin{2^{L_1}\mathbf{x}},\dots,\cos{2^{L_N}\mathbf{x}},\sin{2^{L_N}\mathbf{x}})
\end{equation}
, where $L_i$ represents the $i$-th level frequency function. We apply this function to each of the coordinate values in $\mathbf{x}$ separately, and then concatenate the result in dimension. Elements in $PE(\cdot)$ with different frequencies could handle non-uniform spatial distribution patterns. For instance, low-frequent elements could capture features on a large geographic scale, such as the first-order propagation effects, whereas high-frequent element could focus on features on a smaller scale, such as regional differences among epicenters and sites. 
This process enables us to obtain the location embedding  $h_{L}=\mathbf{NN}(PE(\mathbf{x}))$, which is then added to each input site token, resembling the absolute position encoding in classical Transformer. In our experiment, we set $N=d/4$, where $d$ is the dimension of the model. $\mathbf{NN}(\cdot)$
is a 1-layer FCN.

Previous research has demonstrated that $PE(\cdot)$ exhibits a distance preservation property \citep{Mai2021ARO}. This property implies that for any pair of locations $\left(\mathbf{x}_m, \mathbf{x}_n\right)$, the similarity between them decreases monotonically as the distance $\left| \mathbf{x}_m - \mathbf{x}_n \right|$ increases.
Furthermore, within the attention mechanism in Transformer model, the inner product of the query $\boldsymbol{q}_m$ and key $\boldsymbol{k}_n$ facilitates knowledge transfer between tokens located at $\mathbf{x}_m$ and $\mathbf{x}_n$. However, as noted by \cite{yan2019tener}, the attention mechanism can lead to a reduction in distance-awareness. Relative positional patterns, particularly distance, are critical for modeling the first order propagation effects.
To address this challenges, we introduce Rotary Position Embedding (RoPE) as proposed by \cite{su2024roformer}, which explicitly incorporates relative position dependencies into the self-attention mechanism using a rotation matrix derived from absolute positions. In our approach, we rotate the query $\boldsymbol{q}_m$ and key $\boldsymbol{k}_n$ by angles proportional to their respective locations, and obtain the updated query $\boldsymbol{q}^*_m=\boldsymbol{R}^d_{\Theta,m}\boldsymbol{q}_m$ and the updated key $\boldsymbol{k}^*_n=\boldsymbol{R}^d_{\Theta,n}\boldsymbol{k}_n$, where $\Theta$ are a set of pre-defined parameters indicating the rates of rotation, and $\boldsymbol{R}^d_{\Theta,i}$ is the rotary matrix for token located at $\mathbf{x}_i$. Additional details on RoPE can be found in the Appendix \ref{appendix: RoPE}.
Similar to the work by \cite{vaswani2017attention}, we define $\Theta=\left\{ \theta_i = b_s^{-4(i-1) / d}, i \in [1, 2, \ldots, d / 4] \right)$. To maintain the distance preservation property, we strive for the inner product to decay as the relative position increases, a principle referred to as the long-term decay property in Transformer models.
Experimental results indicate that this long-term decay property is closely related to the selection of $b_s$. Unlike the discrete position indices typically employed in Transformer models, geographic locations are continuous, suggesting that the choice of $b_s$ may need to be adapted from the conventional values used in standard Transformers. For a comprehensive comparison of different positional encoding methods and pre-defined parameters, please refer to Section \ref{location embedding comparison}. 

\subsubsection{Masking Strategy}
\label{section:masking strategy}
In this research, we have established two spatial masks and one temporal mask to mask the stations ($M_S$), the waveforms of the stations ($M_{WS}$), and the evolution of the waveforms over time ($M_{WT}$). Different spatiotemporal masks correspond to three different seismic GMP tasks (Figure\ref{fig:model_arch}). 

\paragraph{Station Mask ($M_S$ )} is applied to the intensity embeddings. When a site is masked by the station mask $M_S$, it indicates that the intensity observation at this site is not available. During the training of the forecasting-interpolation model, intensity embeddings for each event are masked by a variable station mask rate $r_s\in[0,1]$.  When $r_s=1$, all intensity tokens for that event are masked, aligning with the forecasting task, while $r_s<1$  corresponds to the interpolation task. $r_s$ is sampled from a truncated $\arccos$ distribution with density function $p(r)=\frac{2}{\pi}(1-r^2)^{-\frac{1}{2}}$ \citep{chang2023muse}. The expectation for this distribution is 0.64, reflecting a strong bias towards higher masking rates. It helps the model to learn the difficult forecasting task. The prediction loss is computed solely on the masked stations. It is important to note that all intensity embeddings must be masked in the EEW task, as complete waveform information for incoming earthquakes is not available. During the inference phase, the intensity embeddings of the target sites are also set to masked. 

 \paragraph{Wave-Station Mask ($M_{WS}$) } is applied to the waveform embeddings when the waveforms are not available.  A wave masking rate $r_w$ is sampled from a uniform distribution. For the interpolation task, waveforms serve as reference information to enhance GMP performance. For simplicity, only site tokens with unmasked intensity embeddings are allowed to have unmasked waveform embeddings, since the intensity could be directly derived from the whole waveform. 
 In contrast, point-based EEW relies on the waveforms from stations closest to the epicenter, in order to infer the source properties. Therefore, during training, the wave embeddings from the $K$ nearest stations to the epicenter should remain unmasked in EEW. Waveform embeddings from all other site token are masked according to rate $r_w$.
 
\paragraph{Wave-Time Mask ($M_{WT}$)} is applied to waveforms prior to encoding them into $h_W$ for real-time purpose in EEW. During the training procedure, we uniformly select a time $t$ (the duration following P-wave arrival) that ranges from 0 s to maximum 17 s for each event. All waveforms data beyond this selected $t$ are set to be zero, effectively masking out the later portions of the waveforms . In inference procedure, the EEW system activates upon detection of the first station triggered by the P-wave and processes real-time data up until the observed time  $t$.

\paragraph{Attention Mask} is utilized to regulate which tokens can attend to one another in the self-attention mechanism. Input site tokens are categorized into two classes: observed and unobserved. For the forecasting-interpolation model, this classification is based on the station mask $M_S$, where intensity embeddings that are not masked by $M_S$ are considered as observed tokens; while for EEW models, this classification relies on the wave-station mask $M_{WS}$. 
Assumes that there are 2 observed tokens ($\textbf{h}_{S,1}$, $\textbf{h}_{S,2}$) and 2 unobserved tokens ($\textbf{h}_{S,3}$, $\textbf{h}_{S,4}$). The model's output should remain invariant to the presence of unobserved tokens, that is, the output of ($\textbf{h}_{S,1}$, $\textbf{h}_{S,2}$, $\textbf{h}_{S,3}$) and ($\textbf{h}_{S,1}$, $\textbf{h}_{S,2}$, $\textbf{h}_{S,4}$) should be identical. Within each Transformer layer, multiple self-attention heads are employed to aggregate the output vectors in the previous layer. For the unobserved tokens,  theirs embedding vectors should be excluded from the self-attention calculations, ensuring that they do not influence the representations of other tokens (see Figure \ref{fig:model_arch} (d)). This mechanism allows the model to focus exclusively on the available data, enhancing its predictive accuracy. 

\subsection{Training and Prediction} 
A basic interpolation model is pre-trained using the complete waveform data. During the pre-training stage, the model is supervised to predict the masked intensities. By adjusting the station mask ratio  $r_s$, this model could be utilized for both ground motion interpolation task (when $r_s<$1) and forecasting tasks (when $r_s=$1). We refer this model as \emph{GMP-uniform}.

For EEW task, we implement two training strategies: 1) Directly train the EEW model using the QuakeFormer architecture, which we refer to as \emph{EEW-single}; 2) Fine-tune the \emph{GMP-uniform}  to develop a model for EEW purpose, referred as \emph{EEW-finetune}. Both models are trained on the real-time waveform dataset using Wave-Time Mask $M_{WT}$.

In the experiment, we employ a 5-layer Transformer with a hidden size of 384, and 8 attention heads, resulting in approximately 2M parameters, which is relatively modest in size. The maximum length of input tokens is set to 50. Adam with $\beta_1=$0.9, $\beta_2=$0.999 is used for optimization. The initial learning rate is 2e-4, and is decreased every 5 epochs by $\gamma=$0.6. We use a batch size of 32, and train the model for 60 epochs. And for fintuning, the initial learning rate is reduced to 5e-5. 

To evaluate the overall performance of our model and estimate its errors, we split our dataset both event-wise and station-wise. First, we separate train events($T_M$), validate events and test events($T_R$) based on the time. Specifically, events occurring 2023 are designated as the test events, while historical events before 2023 are split to train events and validation events in an 85:15 ratio. Records associated with a particular event are assigned to the same subset. Then, we randomly separate reference points into train locations ($S_M$), validate locations and test locations($S_R$) based on the space distribution, following an 80:10:10 ratio.  Records belonging to $T_M+S_M$ are used for model training, while validate events and validate stations are for model selection. We assume that sites in $S_M$ represent seismic stations capable of observing seismic records; consequently, the intensity and waveform embeddings for sites in validation/test set will be all masked by $M_S$ and $M_{WS}$, respectively. Note that, not all stations in $S_M$ will have observations for a given event. During training, implicit station characteristics are modeled using the historical recordings. Model performance assessments are conducted on both $T_R+S_M$ and $T_R+S_R$ sets. We primarily utilize $T_R+S_M$ to evaluate on-site performance for future events, which is critical for the ground motion forecasting task and on-site EEW applications. 
Moreover, accurate shaking estimates at sites lacking historical recordings are essential in real-world scenarios, particularly for tasks such as interpolation for emergency rescue and regional EEW applications. Therefore, we focus on assessing model performance for spatial distribution predictions on $T_R+S_R$. 


\section{Experiment}
We demonstrate that QuakeFormer is an effective uniform architecture for GMP tasks (forecasting, interpolation and early warning task) through comparison with the traditional approaches in these three tasks. The model successfully tackles multiple GMP tasks without modifying the architecture and the training procedure, while also outperforming specialized architectures. We choose the coefficient of determination ($R^2$) as the comparing metric, which represents the proportion of variance explained by the model. We set $L_1$=-2, $L_N$=5 and $b_s$=5000 in the experiment. 

\paragraph{Baseline} For the forecasting comparison, we select ASK-14\citep{Abrahamson2014ASK14}, which is a widely used GMPE for probability seismic hazard analyses. For interpolation, we compare our results with the Conditional Multivariate normal distribution approach (MVN) \citep{Worden2018interpolationMVN}, which serves as the core algorithm in ShakeMap. In MVN approach, ground motion forecasting models provide the initial estimates of ground motion, then the estimations at target locations are conditioned upon the recorded ground motions at observed locations. To evaluate the effects of interpolation exclusively, we utilize QuakeFormer as the forecasting model in our experiments. 
We compare QuakeFormer with two widely used earthquake early warning algorithms: the Estimated  Point Source (EPS) and the Propagation of Local Undamped Motion (PLUM), based on point-source estimation and  wave propagation, relatively. 
EPS, such as ElarmS and ShakeAlert systems, first generates a point-source estimate, which includes maginitude and location, then the spatial ground shaking is calculated using attenuation relations in forecasting models.
We use catalog hypocenters and magnitude reports provided by ShakeAlert as source parameters. GMPE (ASK14) and the interpolation approach (MVN) are subsequently applied to obtain the regional PGA distribution
In PLUM method\citep{kodera2018propagation}, the predicted seismic intensity at a target site is taken to be the maximum of the observed real-time intensities in a circular region of radius $r$. Details about these baseline models could be seen in the Appendix.


\subsection{Ground motion forecasting and interpolation}
\begin{figure}[h]
    \centering
    \includegraphics[width=\linewidth]{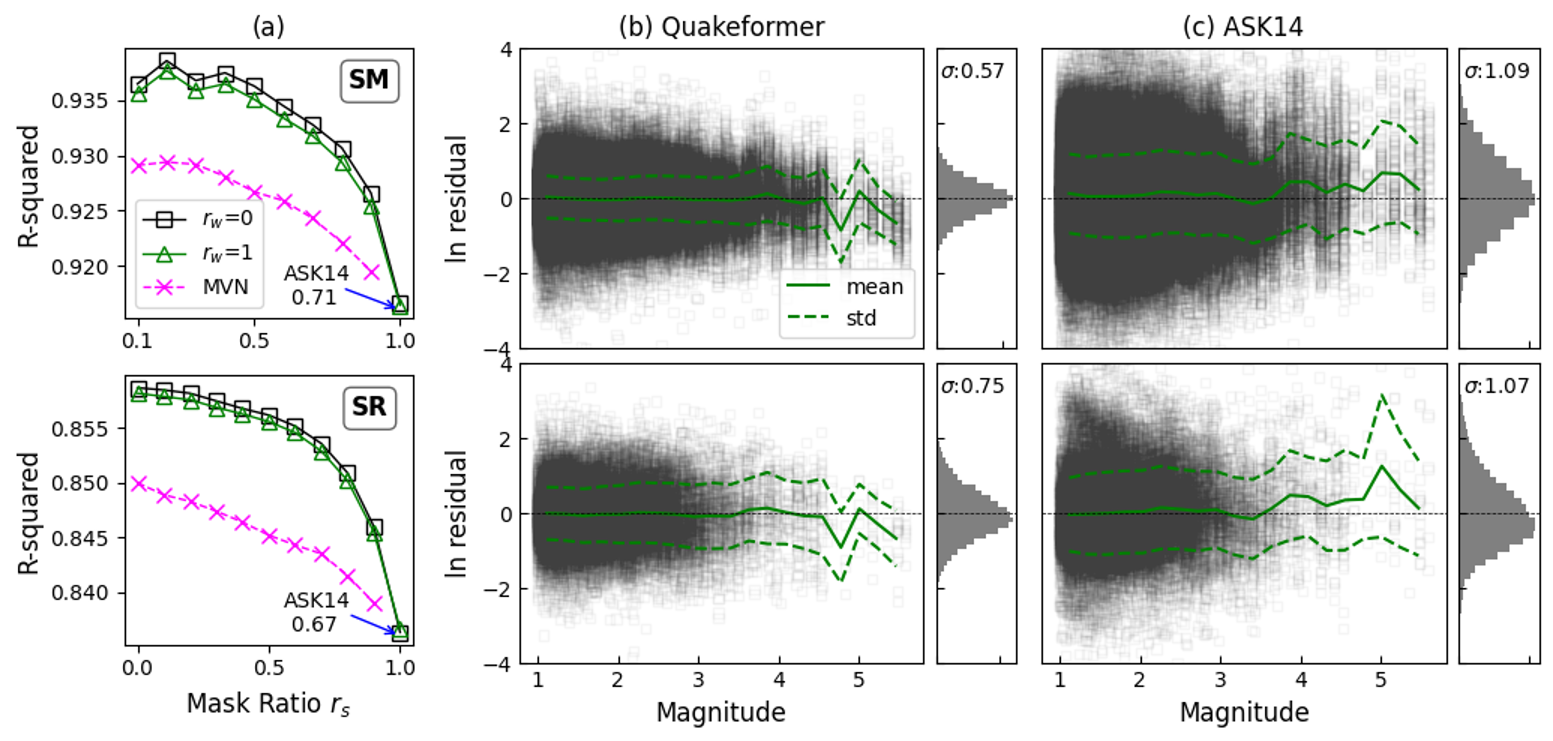}
    \caption{GMP forecasting and interpolation model comparing: (a) The $R^2$ scores for QuakeFormer, ASK14 (forecasting) and MVN approach (interpolation) across different station mask ratio $r_s$ and wave mask ratio $r_w$ on $T_R$. (b) if the total residual distributions for QuakeFormer ($r_s$=1) and (c) is for ASK14. For each column, the upper are results on TR+SM set, while the lower are for TR+SR set. Every point refers to a recording. The green solid line represents the mean, and dashed lines represent the standard derivation around the mean.} 
    \label{fig:interp-r2-residual}
\end{figure}

Figure \ref{fig:interp-r2-residual}(a) demonstrates that  QuakeFormer outperforms MVN and ASK14 in terms of prediction performance. Moreover, QuakeFormer provides consistent prediction results with across varying numbers of input stations. As the station mask ratio $r_s$ decreases, the prediction task shifts from forecasting to interpolation. We observe a significant improvement in performance when the masking ratio shifts from 1 to 0.8, indicating that even a limited number of observations can aid in capturing the propagation patterns of a specific earthquake, thereby reducing the uncertainty associated with predictions.  What's more, comparing with $r_w=$0 and $r_w$=1, the additional utilization of waveform data also contribute to improve the accuracy of the model, since the recorded waveform contain frequency-based site characteristics. Notably, the $R^2$ scores for seen stations (SM) surpass that for unseen stations (SR), indicating QuakeFormer's potential for excellent performance in regions with historical data. 

Figure \ref{fig:interp-r2-residual}(b) and (c) shows the total residual ($\delta_{ij}=\ln{pred}-\ln{true}$) distributions for QuakeFormer ($r_s$=1) and ASK14, respectively. $\sigma$ is the standard derivation of the total residual. In QuakeFormer, the mean and standard derivation of residuals shows minor magnitude-dependent trends. This demonstrates that the magnitude scaling of ground motions in the whole California is well captured by QuakeFormer.

\begin{figure}
    \centering
    \begin{subfigure}[b]{\textwidth}
        \includegraphics[width=\linewidth]{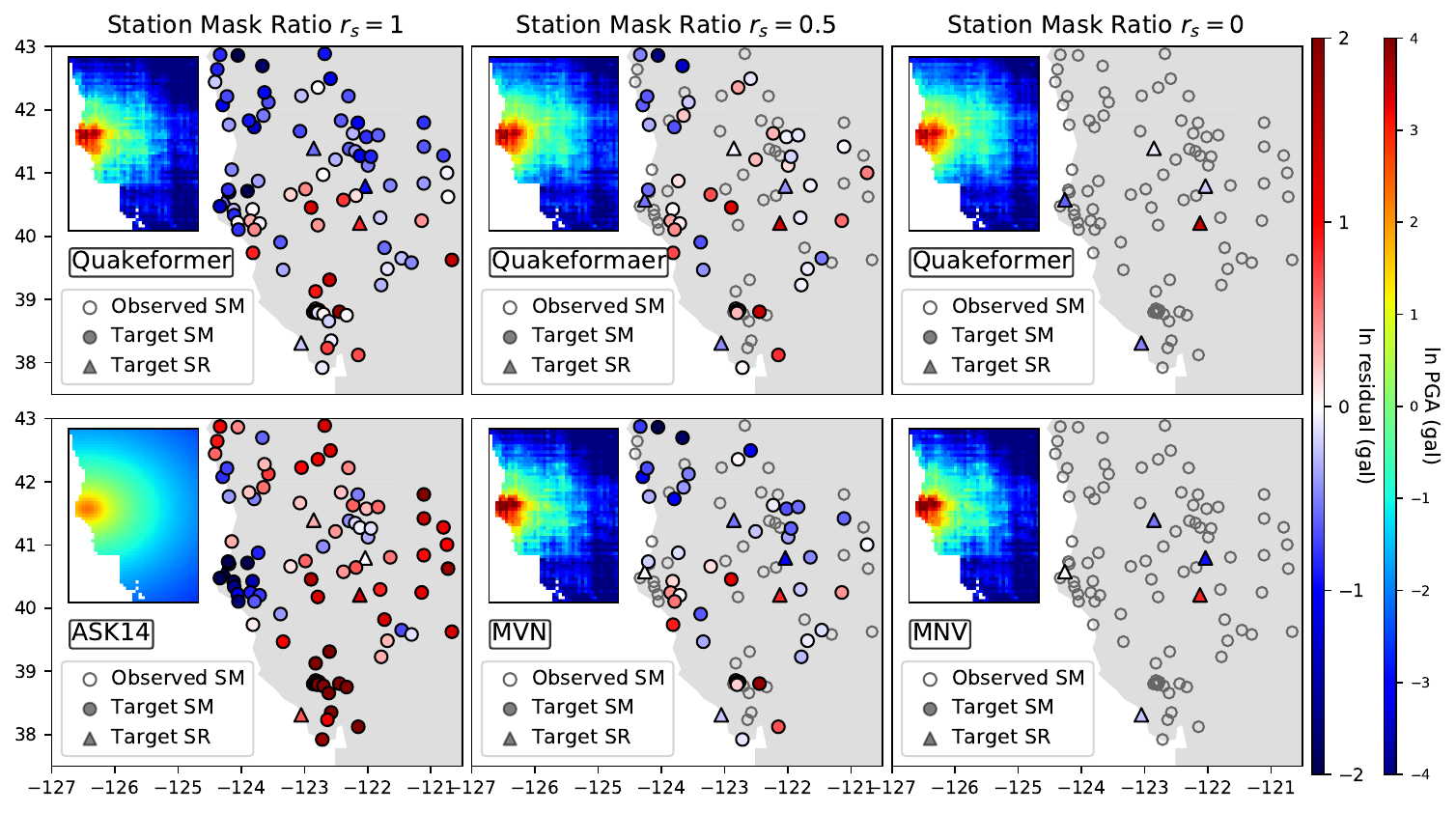}
        \caption{}
    \end{subfigure}
    \hfill
    \centering
    \begin{subfigure}[b]{\textwidth}
        \includegraphics[width=\linewidth]{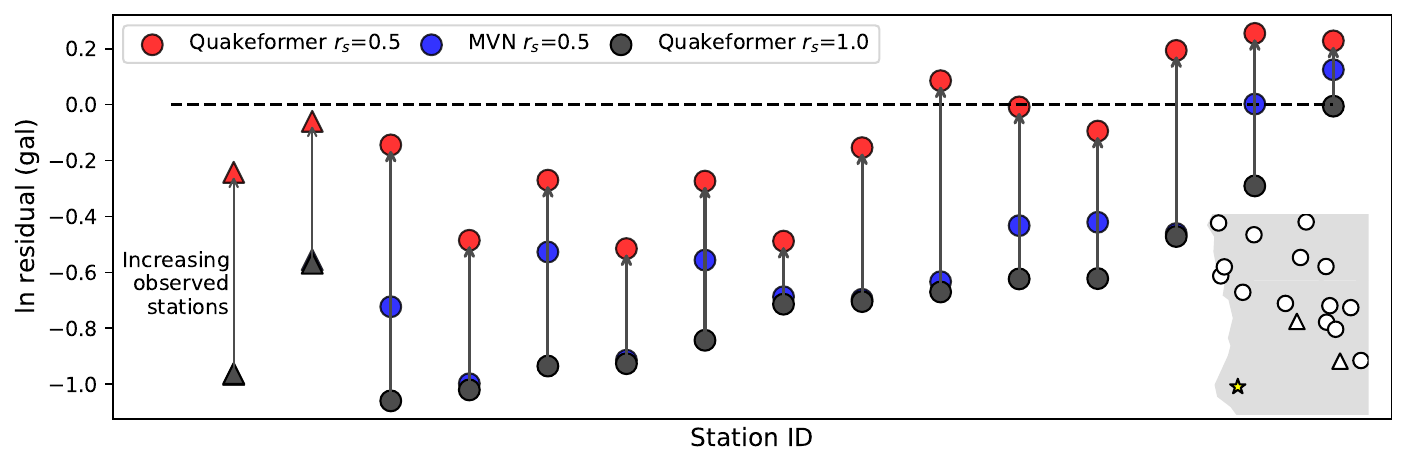}
        \caption{}
    \end{subfigure}
    \caption{Example result (id: nc73827571) for ground motion forecasting and interpolation, compared with QuakeFormer, MVN, and ASK14: (a) displays the predicted ln PGA distributions and the corresponding residuals $\delta$. Solid dots indicate target sites, color-coded based on the residuals $\delta$; while hollow dots represent observed stations. In the top left corner is the map showing the predicted PGA. Moving from left ($r_s=$1, forecasting) to right ($r_s=$0), the number of observed stations (belong to SM) increases from 0 to all. The wave mask ratio $r_w$ is 100\% for all examples presented here. (b) compares the PGA residuals in the northeast region as $r_s$ varies from 1 to 0.5. Each column corresponds to a target site in (a) (when $r_s=$0.5). In the lower right corner, the yellow star denotes the epicenter; circles and triangles represent target sites from SM and SR, respectively. All results are presented in natural log units (gal).  }
    \label{fig:interp_case}
    
\end{figure}

Figure \ref{fig:interp_case} shows an earthquake happened in 11 May 11 2023 with $\textbf{M}$ 5.35 (id: nc73827571), located 15 km southeast of Rio Dell. We employ QuakeFormer, ASK14 and MVN approaches to predict the PGA distribution. Both QuakeFormer and MVN approaches show a reduction in prediction error as the number of observed stations increases, with $r_s$ varying from 1 to 0. 
We notice significantly difference in the propagation effects among these three approaches, particularly in the northeast region. 
Figure \ref{fig:interp_case} (b) demonstrates that QuakeFormer revises the attenuation relation to be weaker, leading to decreased prediction residuals in this area. 
Nevertheless, MVN approach makes only minor adjustment to the spatial trend. This limited responsiveness may stem from the lack of revisions to variances and covariances that account for the spatial averaging of ground motion \citep{Worden2018interpolationMVN}. 
The prediction from ASK14 appears overly smoothed due to the absence of source and site characteristics. Also this simplified version of ASK14 tends to underestimate PGA distribution near the source while overestimates it at distance greater than 100 km, which aligns with the PGA regression curve in the \href{}{\href{https://earthquake.usgs.gov/earthquakes/eventpage/nc73827571/shakemap/intensity}{ShakeMap}}.  More examples are in the Appendix.

\subsection{Earthquake early warning}

\begin{figure}
    \centering
    \includegraphics[width=\linewidth]{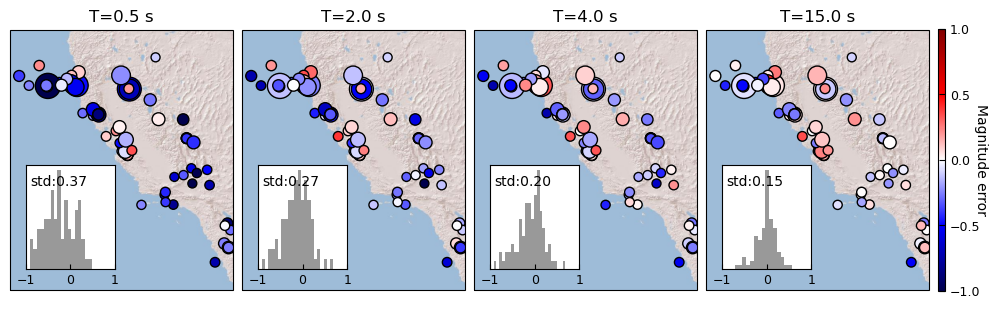}
    \caption{Magnitude evolution by P-wave arrival time ($\textbf{M}>3.5$, 2023)}. We use $\text{EEW-finetune}$ with wave masking ratio $r_w=$0.
    \label{fig:Magnitude distribution by time}
\end{figure}

To evaluate the performance of the EEW task, we conduct tests on 88 relatively large events in 2023 ($\textbf{M}>$3.5). We first compare the source parameters (epicenters and magnitudes), as shown in Figure \ref{fig:EEW-hist-mag-loc}. QuakeFormer demonstrates a more robust estimation of magnitude within 4 seconds after the P-wave arrival, exhibiting a median error of 0.16 magnitude units (standard deviation: 0.2 magnitude units), see Figure \ref{fig:Magnitude distribution by time}, in comparison to ElarmS-3 reports a median magnitude error of 0.3 \citep{chung2019optimizing}. 
However, QuakeFormer shows a slight performance decline compared to ElarmS-3 in earthquake location prediction, yielding a median distance error of 6.81 (standard deviation: 8.81), whereas ElarmS-3 achieves 2.78 km (in-network). 

\begin{figure}
    \centering
    \includegraphics[width=\linewidth]{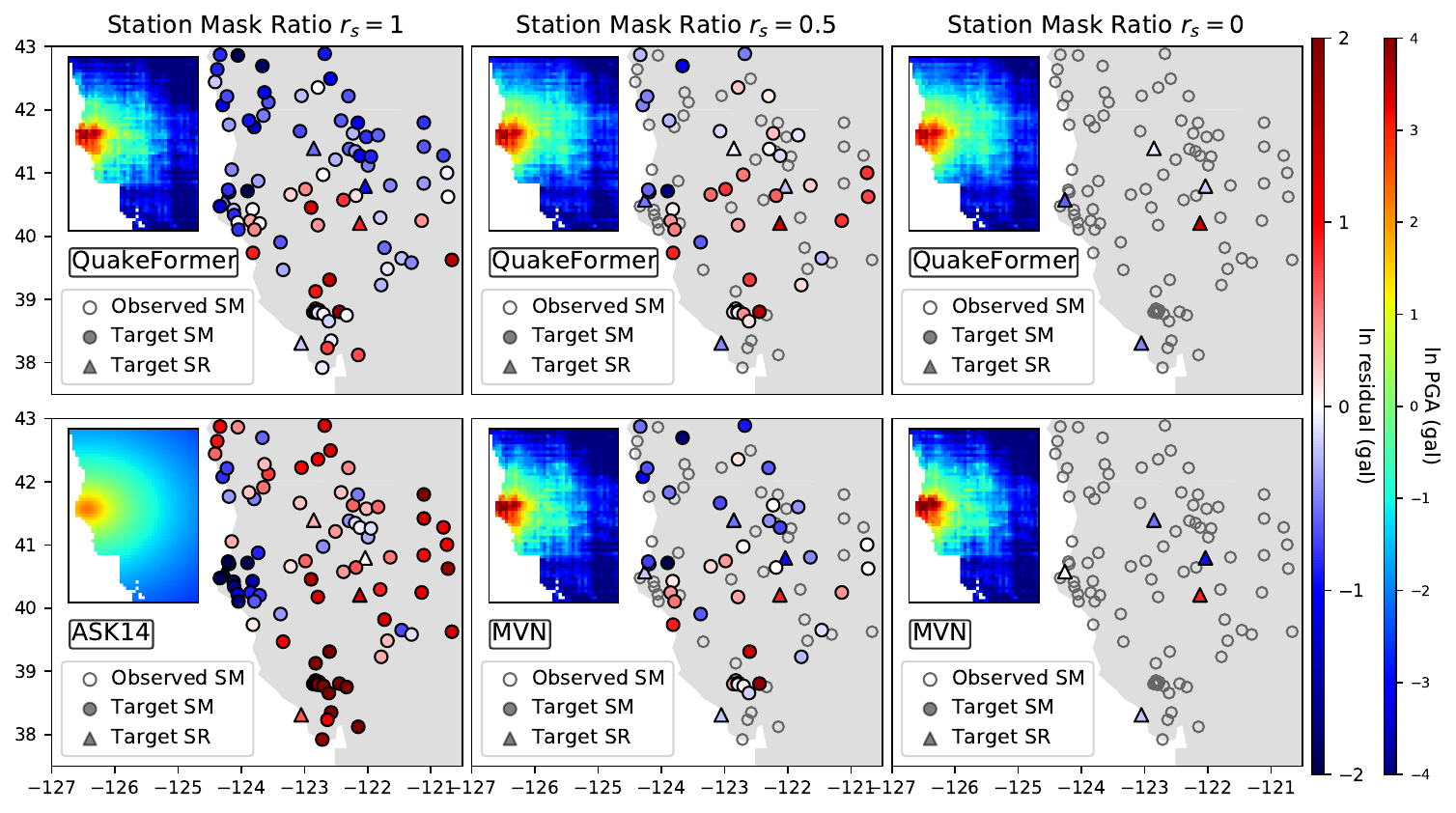}
    \caption{Example result (id: nc73827571, 5.39 $\textbf{M}$) for earthquake early warning, compared with QuakeFormer (\emph{EEW-finetune}, \emph{EEW-single}), EPS and PLUM. The time of P-wave arrival at the nearest station is 5.39 s. At 6 s after the origin time, P-wave arrived at station NC.KMPB, BK.RBOW and NC.KCR. }
    \label{fig:EEW-case}
\end{figure}

Despite QuakeFormer's ability to output source parameters comparable to other EEW systems, its primary advantage in the EEW domain lies in its capability to deliver an end-to-end intensity estimation,  thereby avoiding accumulated errors from the sequential application of source estimation and ground motion forecasting algorithms (always GMPEs).  Figure \ref{fig:EEW-case} illustrates the evolutionary estimation results for a real earthquake (id: nc73827571). At 6 s after origin time (OT), both the QuakeFormer \emph{EEW-finetune} and \emph{EEW-single} models output a rough yet rapid estimation about magnitude, location and PGA distribution. Despite QuakeFormer underestimates the shaking intensity, it's still the quickest approaches that accurately detect the extent of the shaking among the three. 
At 8 s after OT, ShakeAlert reports its first alert based on EPS with magnitude 5.3, which is close to the catalog magnitude 5.39. However, due to the complex source characteristics and the oversimplified nature of GMPEs, the EPS approach significantly underestimates the shaking near the source. 
By 12.6 s after OT, ShakeAlert reports a peak estimation with magnitude 5.5. At this point, 5 stations closest to the source received the S-wave, allowing the interpolation approach (MVN in this analysis) to revise the PGA distribution near the source using the observed real-time intensities. 
PLUM produces accurate shaking estimates near the source; but there is a trade-off between the warning-time and the estimation accuracy\citep{kodera2018propagation}, depending on the selected radius $r$. QuakeFormer exhibits superior performance,  particularly in capturing large-scale propagation effects and fine-grained shaking details . TEAM proposed by \citet{munchmeyer2021transformer} is similar with our \emph{EEW-single} model, as it estimates the shaking directly from the raw waveforms using Transformer architecture. They declares that complex event characteristics, such as stress drop, radiation pattern or directivity, can be implicitly modeled using spatially distributed waveforms, enabling the model to provide a convincing PGA distribution over large regions.  

\begin{figure}
    \centering
    \includegraphics[width=\linewidth]{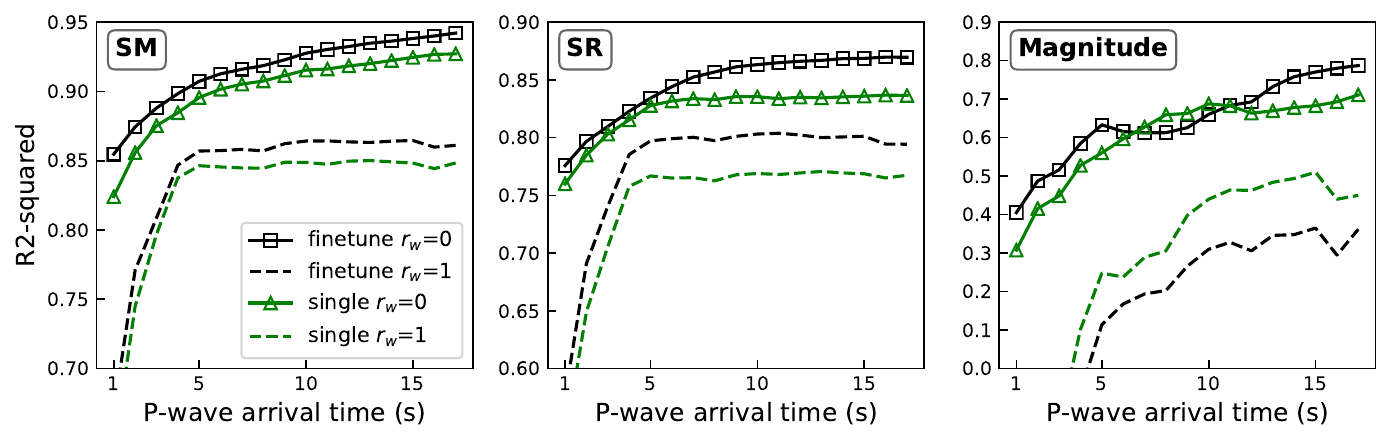}
    \caption{Temporal performance comparison between $\text{EEW-finetune}$ and $\text{EEW-single}$ ($\textbf{M}$, 2023). Curves show the $R^2$ scores for PGA estimates on SM (seen) sites,  SR (unseen) sites, and $\textbf{M}$ after the first station triggered by the P-wave arrival. }
    \label{fig:EEW model compare sinle/finetune}
\end{figure}

We also notice that \emph{EEW-finetune} achieves the smallest prediction error near the source. Figure \ref{fig:EEW model compare sinle/finetune} compares \emph{EEW-single} and \emph{EEW-finuetune} on the large events. The result demonstrates that \emph{EEW-finuetune} outperforms \emph{EEW-single} in both source parameter estimation and shaking prediction.
Note that \emph{GMP-uniform} is trained on the entire historical earthquake dataset, that includes all available stations in space and complete waveforms over time. Complex spectrum characteristics of site responses are extracted and implicitly memorized in the neural network.  Consequently, when finetuned on the \emph{GMP-uniform}, \emph{EEW-finetune} inherits the learned frequency-dependent site effects, enhancing its predictive capabilities in the EEW task.

\begin{table}[]
\caption{Magnitude evolution for earthquake nv73827571}
\begin{tabular*}{\hsize}{@{}@{\extracolsep{\fill}}cccccccc@{}}
\toprule
model                                                                                  & 6 s after OT                                                                          & 8 s atfer OT & 12.6 s atrer OT \\ \midrule
QuakeFormer (finetune)\ & 4.30                                                                                  & 5.57         & 5.51            \\
QuakeFormer (single)   & 3.65                                                                                  & 4.55         & 4.45            \\
EPS                                                                & (only 3 stations triggered)  & 5.3          & 5.5             \\ \bottomrule
\end{tabular*}

\end{table}



\section{Discussion}

\subsection{Comparison of location embedding methods}
\label{location embedding comparison}
We conduct an ablation experiment on the location embedding methods:
\begin{itemize}
    \item M1: use both absolute location embedding ($h_{L}=\mathbf{NN}(PE(\mathbf{x}))$) and relative location embedding (RoPE, $b_s$=5000). $PE(\cdot)$ is based on Eq. \ref{location embedding}, and $L_1$=-2, $L_N$=5. 
    \item M2: only use absolute location embedding, which is same with M1.
    \item M3: only use relative location embedding (RoPE), all location embedding vectors (for events and sites) are replaced by a fixed random vector.
    \item M4: use absolute location embedding ($h_{L}=\mathbf{NN}(\mathbf{x})$) and relative location embedding. The network $\mathbf{NN}$ directly operate on the pair of latitude and longitude.
    \item M5: use absolute location embedding ($h_{L}=PE(\mathbf{x})$) and relative location embedding. Not apply networks on the result of $PE(\cdot)$.
\end{itemize}

\begin{figure}[h]
    \centering
    \includegraphics[width=\textwidth]{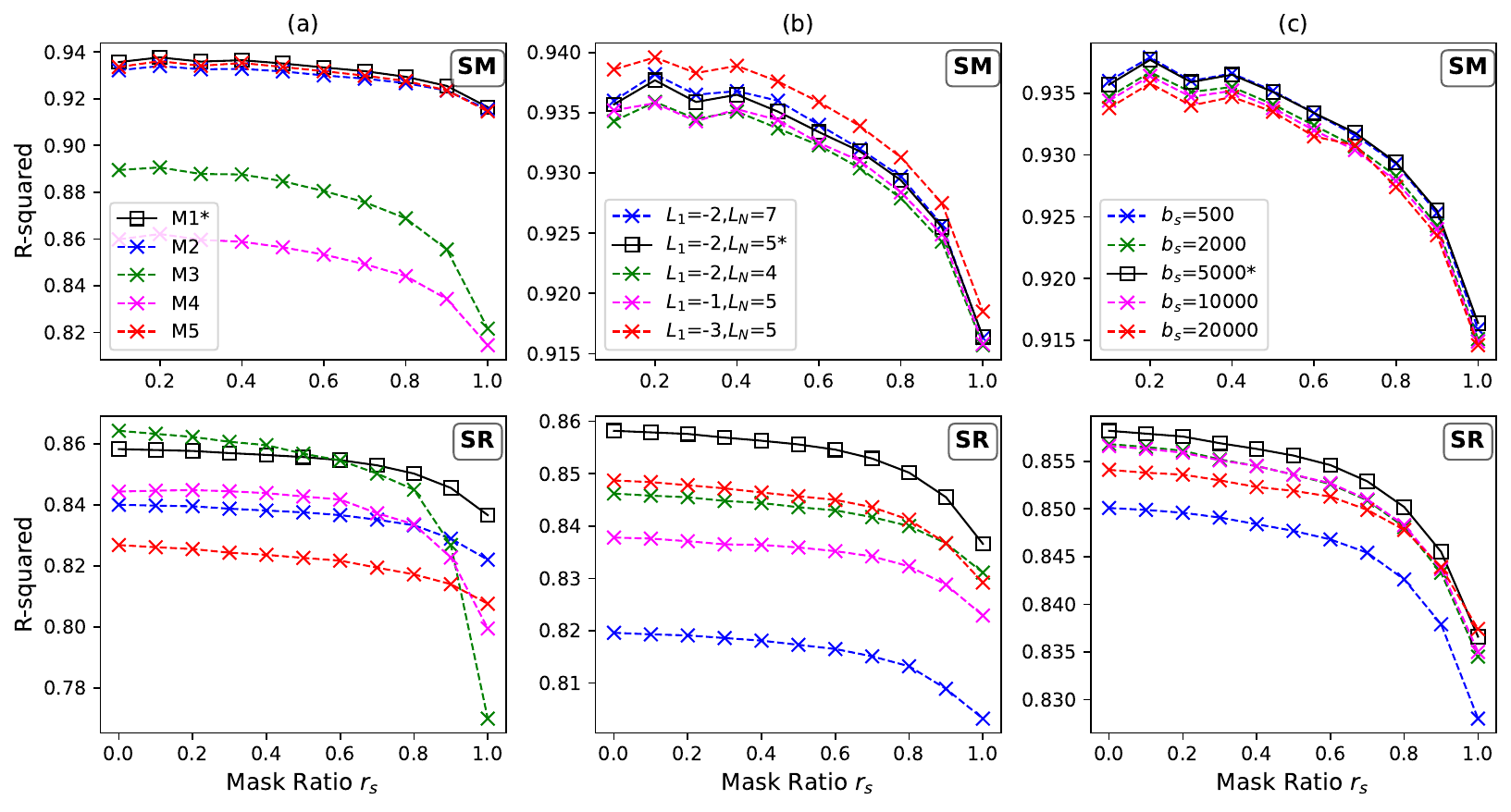}
    \caption{Comparison of different (a) location embedding methods, (b) spatial resolutions in the absolute location embedding ($L_1$,$L_N$), and (c) $b_s$ in relative location embedding. The first line are the $R^2$ scores for PGA predictions from locations with historical recordings in 2023 (TR+SM), and the second line are for locations without historical recordings (TR+SR). Wave ratio $r_w=$1. }
    \label{fig:R2 position embedding}
\end{figure}

Figure \ref{fig:R2 position embedding} (a) shows $R^2$ scores for different position embedding methods, and M1 achieving optimal performance both on SM (seen locations) and SR (unseen locations).  
Deep network tends to learn lower frequency functions \citet{rahaman2019spectral}. Comparing with M4, the result demonstrates that mapping low-dimensional coordinates into high-dimensional space helps downstream network to better fit the ground motion data with high frequency variation, which is consistent with the work by \citet{Mildenhall2020NeRF}. However, we also find a decrease in prediction performance on SR in M1, M2 and M5. In scenario with low station density, model with the sensitivity to high-frequency signals may easily capture spatial high-frequency patterns as well as high-frequency noise, leading to overfitting at locations that were not included in the training dataset. 

What's more, M1 outperforms M2 since the relative spatial dependencies in M1 are explicitly incorporated into the Transformer using RoPE. But if we only encode the relative spatial relationships, like M3, the model transforms into an ergodic GMP model. Consequently, the predictive accuracy diminishes notably when the station masking ratio $r_s$ changing from 0.8 to 1.0 (forecasting), primarily due to the absence of regional distinctions. This observation suggests that location-based event effects and site/path effects play substantial roles in the total residual  (see Appendix where we separate the total residual in M3 into event/site/path terms). 
Comparing M1 with M5, we find applying a shallow neural network after $PE(\cdot)$ is essential for our task, as it enhances the model's representative capacity. 

\begin{figure}[h]
    \centering
    \begin{subfigure}[b]{\textwidth}
        \centering
        \includegraphics[width=\textwidth]{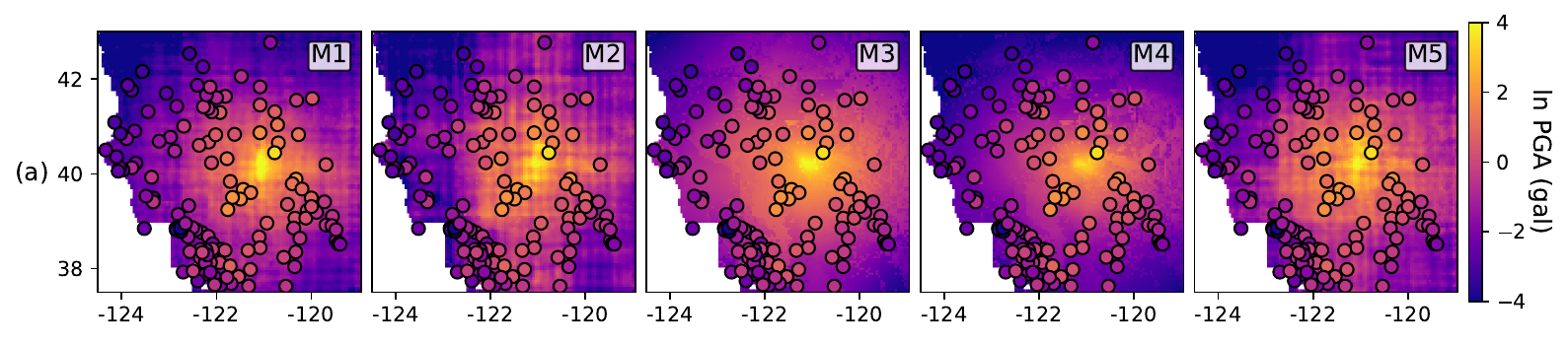}
    \end{subfigure}
    \hfill
    \begin{subfigure}[b]{\textwidth}
        \centering
        \includegraphics[width=\textwidth]{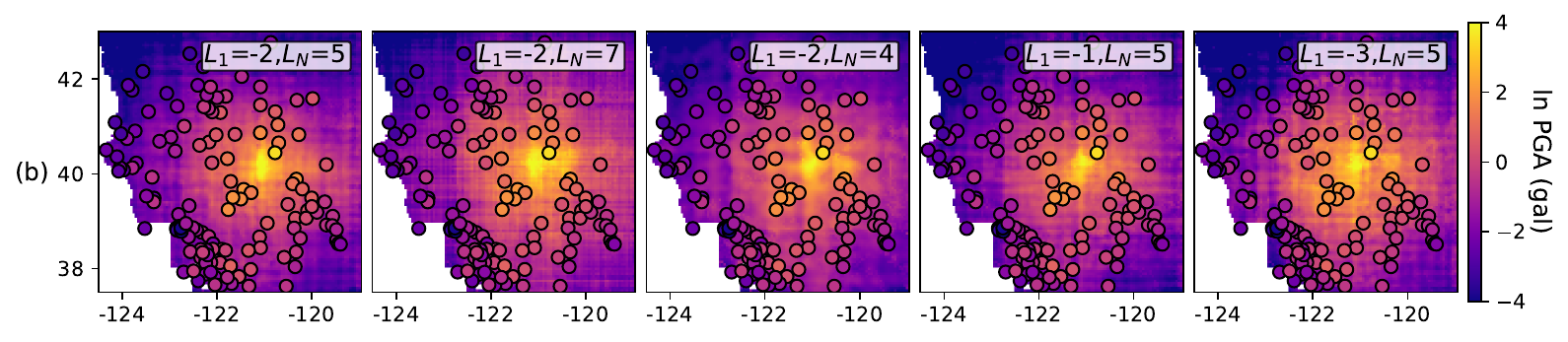}
    \end{subfigure}
    \caption{Comparison of the predicted PGA distributions using different (a) location embedding methods, (b) spatial resolutions in the absolute location embedding ($L_1$,$L_N$). The event id is nc73886731. Station mask ratio $r_s=$1. The scatter points refer to the recorded PGA. Values are given in ln unit.}
    \label{fig:pos embed, gm example}
\end{figure}

We also compare three important pre-defined parameters in the absolute location embedding ($L_1$, $L_N$) and relative location embedding ($b_s$). 

\begin{figure}[h]
    \centering
    \begin{subfigure}[b]{0.49\textwidth}
        \centering
        \includegraphics[width=\textwidth]{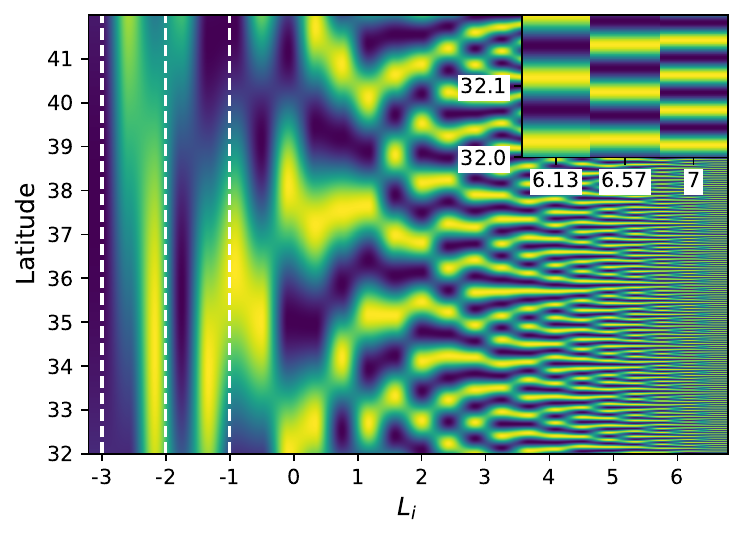}
        \caption{}
        \label{absolute embedding}
    \end{subfigure}
    \begin{subfigure}[b]{0.49\textwidth}
        \centering
        \includegraphics[width=\textwidth]{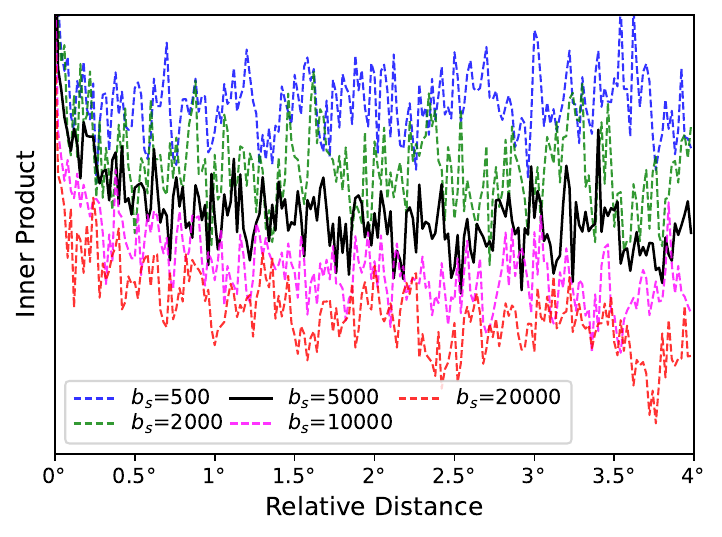}
        \caption{}
        \label{relative embedding}
    \end{subfigure}
    \caption{(a) The partial location embedding vectors generated by Eq. \ref{location embedding}, $\sin{(2^{L_i}\cdot lat)}$.  Each row is a vector at specific latitude. The longitude for these embedding vectors are fixed to 120W. ; (b) Comparing the inner products of updated $\boldsymbol{q}_m^*$ and $\boldsymbol{k}_n^*$ across different rotary base parameter $b_s$ as the relative distance of location $\mathbf{x}_m$ and $\mathbf{x}_n$ increases. }
    \label{fig:location embedding vector}
\end{figure}

Figure \ref{absolute embedding} visualizes the location embedding values as they vary by latitudes across different spatial resolutions in Eq. \ref{location embedding}. In Eq. \ref{location embedding}, $L_1$, $L_N$ reflect the maximum and minimum wavelengths used to decompose the coordinates, respectively. For a spatial reference point $p_i$ at location $\mathbf{x}_i$, the embedding values obtained by the $j$-th level sinusoidal function in Eq. \ref{location embedding} are $\sin{2^{L_j}\mathbf{x}_i}, \cos{2^{L_j}\mathbf{x}_i} \in [0,1]$.  As the value of $L_j$ increases, the wavelength of sinusoidal function decreases, leading to greater spatial resolution. Furthermore, due to the periodicity of sinusoidal functions, using multi-scale sinusoidal $PE(\cdot)$ effectively divides space into "grids" of multi-scale wavelengths, the embedding value represents the relative position to the $j$-th level grid lines.  

When $L_1$ is excessively large, the relative position of $p_i$ within the entire region becomes indistinguishable. Therefore, as depicted in Figure \ref{absolute embedding} when $L_1=$-1, points in different locations yield identical encoding results. It is problematic especially for the locations without historical recordings (see Figure \ref{fig:R2 position embedding} (c) when $L_1=$-1). To better preserve the location distinction across the entire region, we can apply the Nyquist theorem to determine $L_1$. Specifically, the minimum wavelength should be twice the region's scale, expressed as 
\begin{equation}
    L_1<log_2(\pi/R)
\end{equation}
$R$ represents the scale of the entire region. For California, where $R$ is approximately 10 degree, we set $L_1=-2$ for this research. 

Moreover, when comparing $L_N=$5 and $L_N=$7 in Figure \ref{fig:R2 position embedding}, we observe that increasing $L_N$ leads to slightly higher prediction accuracy for SM, but will cause a drastic decrease on SR performance. The reason is, if $L_N$ is overly large, the maximum spatial resolution becomes excessively fine, causing even location embeddings from nearby positions to exhibit minimal similarity. This high-frequency spatial variance aids stations with historical records in capturing location-specific features more effectively, but it may also lead to an overemphasis on the spatial noise.

Figure \ref{fig:R2 position embedding} (c) shows the comparison of different base parameters $b_s$ in RoPE.  To illustrate the correlation between $b_s$ and the distance preservation property, we conduct a simple experiment. Supposing the query $\boldsymbol{q}_m$ and key $\boldsymbol{k}_n$ from location $\mathbf{x}_n$ and $\mathbf{x}_m$ are both all-one vectors. Specifically, we set the coordinate of $\mathbf{x}_n$ to (32N,120W), while varying the coordinates of $\mathbf{x}_m$ from (32N,120W) to (35N,120W). Figure \ref{relative embedding} depicts the inner product (similarity) as the relative distance changes.
It can be observed that when the base parameter $b_s$ is too small, RoPE fails to exhibit long-distance attenuation, leading to oscillations in the attention computation when ROPE is introduced. However, as the value of $b_s$ is too large, the degree of remote attenuation gradually diminishes. Consequently, the predictive performance of QuakeFormer also declines. We also found that the parameters of Rope $b_s$ have a relatively smaller impact on the model's performance, compared to $L_1$ and $L_N$. 

In addition to the statistic metrics, we also compared the estimated PGA distributions across different position embedding approaches and resolutions (Figure \ref{fig:pos embed, gm example}). 
We observe a checkerboard-like pattern indicative of severe spatial overfitting in the predictions, particularly pronounced when  $L_N$ is large.
The distributions estimated by M3 and M4 approaches appears smoother than others, as these two approaches do not utilize absolute location embedding $PE(\cdot)$. This observation underscores that while $PE(\cdot)$ could enhance prediction performance, it may also lead to spatial overfitting, especially if the training sites lack sufficient density, resulting in unnatural PGA distribution on a large scale. 
Comparing M1 and M2, we also notice that the relative location embedding RoPE is helpful to diminish the overfitting pattern. In regions with limited historical recordings, the relative position embedding that models propagation effects should take precedence in the prediction process. Importantly, we did not identify temporal overfitting in our analysis, which suggests that predictions of future ground motion are reliable. 

Therefore, the choice of position embedding methods and parameters should be tailored to the specific real-world applications. If the focus is primarily on on-site predictions, especially for critical fundamental infrastructures such as nuclear stations, or if the studying region has high-density station array, employing QuakeFormer with high-resolution $PE(\cdot)$ may be the most suitable choice to capture the site-specific characteristics. Whereas, if the prediction are intended for regional decision-making, or the station array is sparse, we recommend carefully select the maximum location embedding resolution $L_N$, or even considering not using $PE(\cdot)$.

\subsection{Insights into the nonergodic residual components}
In order to understand how close we are to completely eliminating repeatable effects in the total variability, we decompose the total residuals $\delta_{ij}=\ln{pred}-\ln{true}$ into event and site residuals. 

The total residuals $\delta_{ij}$ for event $i$ and station $j$ could be explained as the combination of between event residuals and with-event residuals. The first term $\delta E_i$ is also referred as event term, representing the residual caused by the remaining difference in the specific source excitation. The last term $\delta W_{ij}$ represents the residual caused by the specific path/site effect, and could be further decomposed into site term $\delta S_j$, remaining aleatory single-station residual $\delta WS_{ij}$.  
To simplify the analysis and avoid the complexities arising from the interplay between between-event residuals and site effects, we compute $\delta E_i$ based on the residuals only from site SM. The standard derivations $\delta E_i$, $\delta L$,  $\delta W_{ij}$ and $\delta S_j$ are $\tau$, $\tau_L$, $\phi$ and $\phi_S$, respectively. 
Notably, location-based event term $\delta L$ and site term $\delta S_j$ are epistemic components that are repeatable and could be modeled. 
More details about the residual decomposition process could be seen in Appendix. Table \ref{table: std comparison} compares the fully nonergodic components identified in other studies in California, revealing significantly smaller values for $\tau_L$ and $\phi_S$ in our study. This result demonstrates that QuakeFormer is good at modeling regional differences. 


\begin{table}[]
\centering
\caption{Comparison of standard derivation estimated in this study and previous studies in California}
\label{table: std comparison}
\scalebox{0.75}{
\begin{tabular}{cccccccccc}

\toprule

    Component & \begin{tabular}[c]{@{}c@{}}This Study PGA\\ TR+SM\end{tabular} & \begin{tabular}[c]{@{}c@{}}This Study PGA\\ TR+SR\end{tabular} & \begin{tabular}[c]{@{}c@{}}\citet{baltay2017uncertainty}\\ PGA\end{tabular} & \begin{tabular}[c]{@{}c@{}}{ \citet{villani2015repeatable}}\\ CS 3.0 s\end{tabular} & \begin{tabular}[c]{@{}c@{}}{ \citet{villani2015repeatable}}\\ CS 3.0 s\end{tabular} \\ \midrule
    $\sigma$  & 0.57                                                           & 0.75                                                           & 0.97                                                 & 0.707                                                     & 0.71                                                      \\
    $\tau$    & 0.34                                                           & $\backslash$                                                           & 0.45                                                 & 0.46                                                      & 0.39                                                      \\
    $\phi$    & 0.48                                                           & 0.61                                                           & 0.86                                                 & 0.53                                                      & 0.59                                                      \\
    $\tau_L$  & 0.17                                                           & $\backslash$                                                           & 0.35                                                 & 0.42                                                      & 0.25                                                      \\
    $\phi_S$  & 0.26                                                           & 0.45                                                           & 0.75                                                 & 0.37                                                      & 0.48                                                      \\
    \bottomrule
\end{tabular}}
\end{table}


\subsubsection{Location-based Event residual}
For moderate-magnitude earthquakes, the event term of high-frequency ground motion, including PGA, is correlated to the level of stress drop \citep{hanks1979b}. Meanwhile, regional differences have been identified in both large and small scales. \citet{baltay2017uncertainty} separated the between event residual $\delta E_i$ into a location-based event term $\delta L$, and the remaining random component $\delta E_i^0$. The result shows that $\delta L$ calculated by earthquake clusters is related to the regional average stress drop, and the estimated stress drop information could be incorporated into GMPE so as to reduce the location-based event residual. 

Although QuakeFormer does not explicitly incorporate prior information about the stress drop into the model, it has the potential to memorizing all the historical information into its neural network and update its interpretation about the source based on the observed records. To demonstrate it, we visually identify 24 earthquake clusters in California (Figure \ref{fig:event cluster in 2023} and Table \ref{cluster number}), that have high-density earthquakes with similar locations and depths. All events in each cluster lie within an area of 2.5 $km^3$. The histogram of $\delta L$ for the 24 clusters could be seen in Figure \ref{fig: event term histogram}. In the ground motion forecasting task, the median absolute $\delta L$ is 0.07, and the std. dev. ($\delta L$)$=\tau_L=$ 0.17 (in ln units).  And for locations with enough historical events (more than 150), the variability of $\delta L$ from these locations  is quite small, the std. dev. ($\delta L$)$=\tau_L=$0.07 (in ln units). We also find a special cluster in North California, namely Cluster 12. At this location, a total of 79 earthquakes occurred in 2023, but no earthquakes had occurred before 2023. Figure \ref{fig:between event residual on cluster 12} shows that QuakeFormer underestimates the event effect of Cluster 12 in the ground motion forecasting task, exhibiting $\delta L=$-0.43. Although no historical earthquakes available in this cluster, QuakeFormer is capable to update the source characteristic  based on the observed records. If we input intensity records from 50$\%$ of seismic stations in this area ($r_s=$0.5), $\delta L$ could be reduced to -0.22 (see Figure \ref{fig:between event residual on cluster 12}).  


\subsubsection{Site terms and $V_{S30}$}
Finally, we discuss the site terms and their correlation with $V_{S30}$, the frequently utilized site effect. 

\begin{figure}[h]
    \centering
     \includegraphics[width=\linewidth]{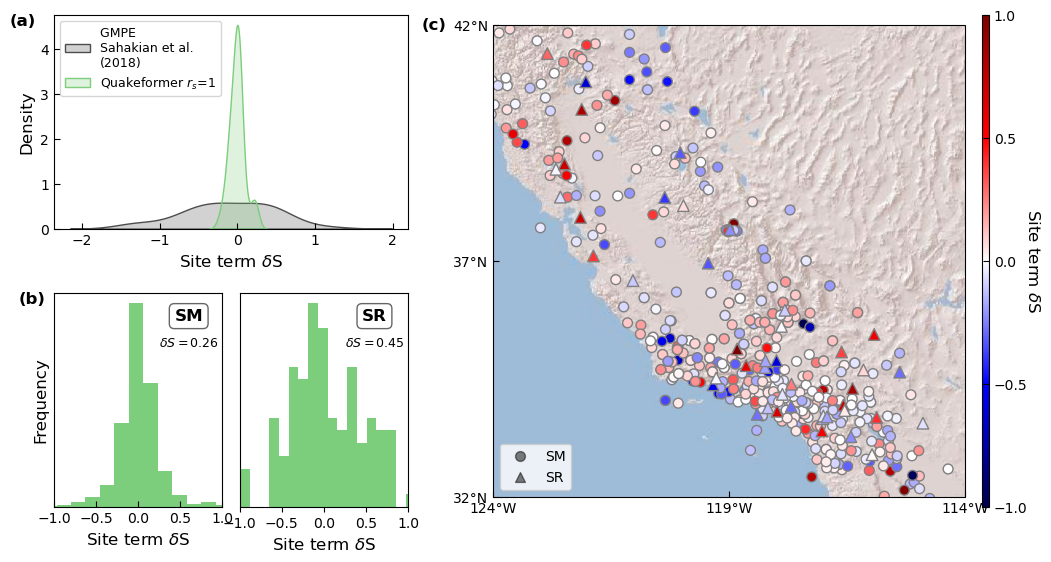}
    \caption{Site term statistic: (a) Kernel density plot for QuakeFormer ($r_s=$1) and the previous GMPE study on Caltech (CI) network. (b) Histograms of site terms on SM and SR sites. (c) Map view of the site terms for stations in the dataset. For comparison, we only compute site terms on stations using high broad band hign-gain seismometers (HH). All results are in ln units.}
    \label{fig:Site term statistic}
\end{figure}

In Figure \ref{fig:Site term statistic} (a) and Table \ref{table: site term comparison}, we compare the  site term $\delta S$ between QuakeFormer and the previous study on GMPE approaches \citep{sahakian2018decomposing}, utilizing data from 38 stations in the Caltech (CI) network in South California.
For the same station, site terms in QuakeFormer are significant smaller, even at locations without historical recordings (SR). 
In QuakeFormer, site specific features are implicitly learned from the  big-volume historical records in our dataset, leading to a substantially reduced site term for SM sites.  
For sites in SR, the improved performance can be attributed to QuakeFormer's enhanced ability to capture location dependencies. This capability allows the model to better interpolate and estimate ground motion characteristics even for regions without extensive data. Figure \ref{fig:Site term statistic} (b) further contrasts the sites terms between SM and SR sites within the dataset, revealing smaller site terms for SM sites. 


Figure \ref{fig:R2 Vs30} presents the prediction and interpolation results under different station masking rates. From this, we observe that the inclusion of the site parameter $V_{S30}$ has almost no impact on the prediction of station locations in the training dataset $S_M$, but it can enhance the prediction in the testing dataset $S_R$. However, even with the incorporation of $V_{S30}$ in the model, the stations in the testing dataset cannot achieve the same level of performance as those in the training dataset. There could be two possible reasons for this: 1) The map of $V_{S30}$ are based on proxy techniques, which may result in ill-described $V_{S30}$; 2) Although $V_{S30}$ is correlated with the deeper velocity structure that reflects the site amplification, it's not the only key factor. The absolute position coordinates we used imply richer site conditions such as topography, soil depths and parameters, etc.. From historical earthquake recordings, the stations in the training set could capture the high-dimensional site features at specific locations, leading to improvement in prediction.

\subsection{Limitations}
In training sites (SM), QuakeFormer effectively captures the intrinsic geographic features, resulting in commendable performance. However, for sites without historical records (SR), we observe that the estimation error remains significant. This discrepancy may arise from the spatial variability of geographic features, which can change at a much higher frequency than the seismic stations. Additionally, existing low-dimensional geographic information, such as $V_{S30}$, does not fully encompass the site-specific characteristics. Consequently, the model faces challenges in accurately interpolating the distribution of geographic features based on the available station observations.  To address this issue, incorporating additional information, such as geological survey data and three-dimensional velocity models, into the deep learning framework could enhance predictive accuracy. Furthermore, the comparison between SM and SR sites also indicates that, deploying stations (which can be mobile stations) in strategically important locations can significantly improve the accuracy of ground shaking predictions for those sites.

We find QuakeFormer may underestimate the shaking intensity near the epicenter. It may due to the imbalance of large-intensity records from near source stations or from large magnitude events. Utilizing synthetic data, such as CyberShake, could enhance model's knowledge of these scenarios and is expected to improve its extrapolation capacity \citep{monterrubio2024machine}. improve its extrapolation capacity , incorporating intensity data from people's feeling may also helps us to know the intensity distribution during big event \citep{atkinson2007did}. Moreover, QuakeFormer exhibits a larger epicenter estimation error compared to estimations from ElarmS-3. The poorer location estimation may stem from the low spatial resolution in the output layer. Decomposing the predicted location multi-scale manner (similar with the position embedding layer we used in the input embedding layer) may improve the epicenter estimation.


\section{Conclusions}

We develop a deep learning based uniform architecture, QuakeFormer, which is capable to be applied to three important ground motion prediction tasks: forecasting, interpolation and early warning. Built upon a simple Transformer framework with absolute and relative position encoding method using the proposed masking strategy, QuakeFormer successfully combines these three tasks into one architecture for the first time, and obtains convincing predictions upon these three tasks. 
QuakeFormer could separately train these three task  (single model), or could jointly train the forecasting and interpolation tasks by using the masking strategy for stations and waveforms (uniform model). Then this basic uniform model could be finetuned to early warning tasks. Also, we demonstrate that uniform model outperforms the single model, especially in the early warning task. Our analysis reveals that it's the ability to integrate the absolute regional differences and the relative spatial dependency that makes QuakeFormer has outstanding performance. In future, we hope to incorporate recent studies on the velocity model, topology and more other geographic information into QuakeFormer, and finally obtain a fully nonergondic ground motion prediction model.

\section{Data and resources}
Waveform data, metadata, or data products for this study were accessed through the Northern California Earthquake Data Center (NCEDC) and Sourth California Earthquake Data Center (SCEDC). The $V_{S30}$ \citep{Eric2022Vs30} and \href{https://apps.nationalmap.gov/downloader/#/}{elevation dataset} are both from U.S. Geological Survey (USGS). The implementation of QuakeFormer will be available after paper reception. 

\clearpage
\bibliography{references}

\appendix

\setcounter{figure}{0}
\renewcommand\thefigure{S\arabic{figure}}

\setcounter{table}{0}
\renewcommand\thetable{S\arabic{table}}

\setcounter{equation}{0}
\renewcommand\theequation{S\arabic{equation}}

\section{Datasets}
\begin{figure}[h]
    \centering
     \includegraphics[width=\linewidth]{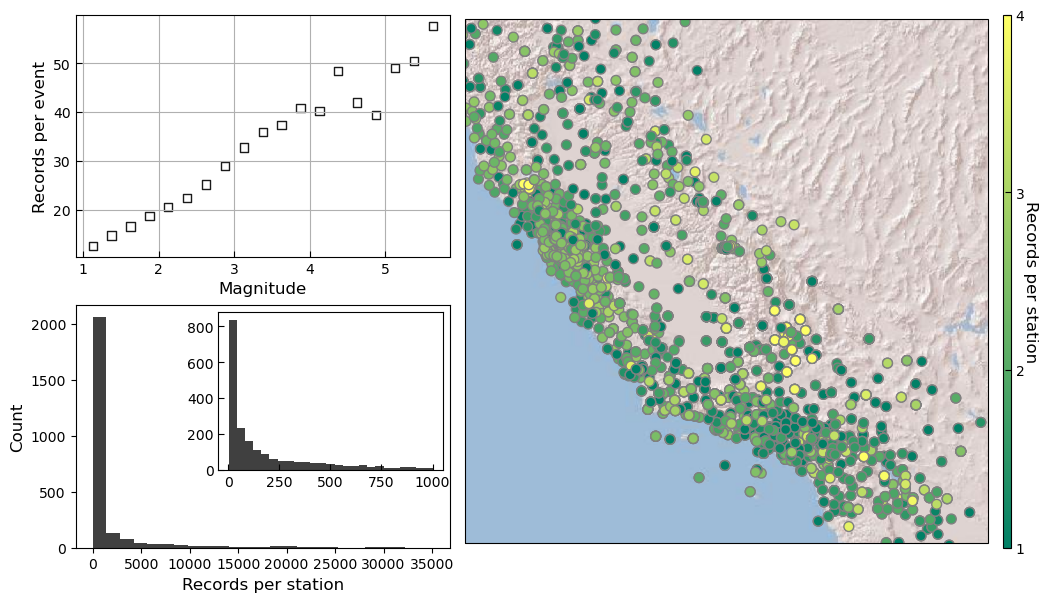}
    \caption{(b) average station records versus $\textbf{M}$; (b) Histograms of the earthquake records per station; (c) Distribution maps of the earthquake records per station.}
    \label{fig:dataset hist}
\end{figure}

Waveforms are aligned and stored in hdf5 format. The standard format of our dataset could be seen in \href{}{\href{https://ai4eps.github.io/homepage/ml4earth/seismic_event_format1/\#folder-structure}{Seismic Event Data (one single file) - AI4EPS} }. The P-phase pick we used is from this dataset. This dataset puts the first P pick of all stations at 30 s and cut a window size of 120 s. In our model, we set $st$, the analysis time before P- wave arrival as 3 s. The intensity label is calculated on the whole duration time $T$, which is calculated by the following criteria: 1) If the trace file contains an S phase,  $T=(t_S-t_P)*3$, where $t_S$ and $t_P$ are the time S-phase and P-phase arrived at this station. 2) If the P phase of the next event  precedes the arrival of S phase, the $T=(t_{P^{'}}-t_P)$, where $t_{P^{'}}$ is the P phase arrival time of the next event. 3) If the trace file does not contain the S phase or the P phase of the next event, similar with \citet{munchmeyer2021transformer}, we use a simplified $T=t_P-Dist/8*f+Dist/2.8*f+10f$, where $Dist$ is epicenter distance and $f$ is the sampling rate. The second term is the approximate P travel time from the source, and the third term approximates the Rayleigh wave speed. Note that, different waveforms recorded by different seismometers at the same location may exist. Therefore, when dividing the stations into training (SM), validation, and training sets (SR), we partition them based on spatial coordinates. Records with different names but located at the same place are assigned to the same station set. During training, we set the maximum length of input site tokens to 50 according to Figure \ref{fig:dataset hist}.

\begin{figure}[h]
    \centering
    \includegraphics[width=\textwidth]{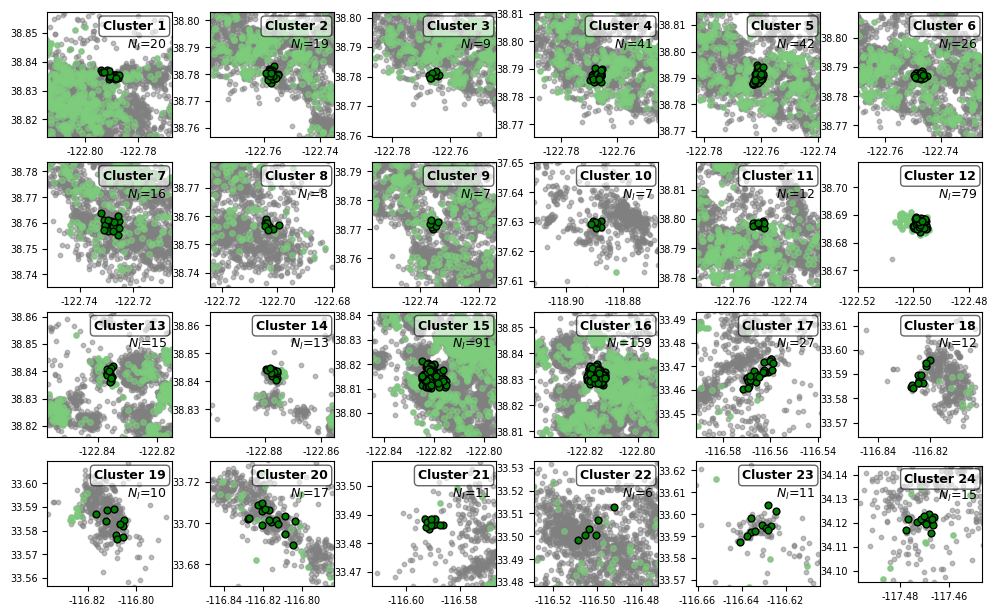}
            \label{cluster-visual}
    \caption{Earthquake cluster: Light gray dots show all seismicity in out study; light green dots shows the earthquakes in 2023;  darker green circles show the earthquake clusters in 2023 we identified for analysis.}
    \label{fig:event cluster in 2023}
\end{figure}

\begin{table}[]
\caption{Number of earthquake events in clusters before 2023 and in 2023}
\label{cluster number}
\begin{tabular}{@{}l|ll|l|ll|l|ll|l|ll|l|ll@{}}

\toprule
ID & 2023 & before & ID & 2023 & before & ID & 2023 & before & ID & 2023 & before & ID & 2023 & before \\ \midrule
1  & 20   &        134& 6  & 26   &        35& 11 & 12   &        15& 16 & 159  &        320& 21 & 11   &        0\\
2  & 19   &        22& 7  & 16   &        134& 12 & 79   &        0& 17 & 27   &        56& 22 & 6    &        118\\
3  & 9    &        11& 8  & 8    &        17& 13 & 15   &        59& 18 & 12   &        1& 23 & 11   &        23\\
4  & 41   &        113& 9  & 7    &        2& 14 & 13   &        45& 19 & 10   &        71& 24 & 15   &        8\\
5  & 42   &        170& 10 & 7    &        2& 15 & 91   &        1004& 20 & 17   &        179&    &      &        \\ \bottomrule
\end{tabular}
\end{table}

\section{Model}
\subsection{Feature encoder} \label{appendix:feature encoder} 
\begin{itemize}
    \item Locations: for sites and epicenter locations, we first apply the multi-scale sinusoidal function $PE(\cdot)$ described in Eq. \ref{location embedding} to map latitude and longitude to high-dimension space. In $PE(\cdot)$, we set the number of scale levels $N=$24, with a output dimension of 96, equal to our model's dimension $D$. Then we apply a 1-layer FCN to obtain the final location embedding. 
    \item Waveforms: We firstly normalized all components of the wave trace by its peak value observed so far. We project the peak value to dimension $D$ with FCN. Convolutional neural network (CNN) is applied to the normalized waveforms to extract frequency features. We add the embedding of peak value and normalized waveforms to obtain the final waveform embeddings.
    \item Attributes: in this experiment, we only include $V_{S30}$, type of seismogram, and site elevation into the site attributes. Every attribute value are separately projected to dimension $D$ using FCN. Then the embeddings are added into the final attribute embedding vectors. More correlated site attributes are allowed to be incorporated into this model.
    \item Intensity: the intensity is the PGA during the whole duration times. We also project it int intensity embedding with dimension $D$ using FCN. 
We use activation function Relu for all layers.
\end{itemize} 

\subsection{Rotary Position Embedding (RoPE)}\label{appendix: RoPE} 
The function formulating the inner product of query $\boldsymbol{q}_m=f_q(\boldsymbol{h}_m, \mathbf{x}_m)$ and key $\boldsymbol{k}_n=f_k(\boldsymbol{h}_n, \mathbf{x}_n)$ is denoted as $g$. In order to incorporate relative position information, the token embeddings $\boldsymbol{h}_m$ and $\boldsymbol{h}_n$, and their relative position $\mathbf{x}_m-\mathbf{x}_n$ are taken as input variables. In other words, $\left\langle f_q(\boldsymbol{h}_m, \mathbf{x}_m),f_k(\boldsymbol{h}_n, \mathbf{x}_n)\right\rangle=g\left(\boldsymbol{h}_m, \boldsymbol{h}_n, \mathbf{x}_m-\mathbf{x}_n\right)$. \citet{su2024roformer} derived a simple formulation, RoPE, which enables preserve relation position dependency. Here we make some adjustment on the RoPE form so that it can be applied to continuous 2D coordinates.
\begin{itemize}
    \item Step 1: Predefine the rotation frequency  $\Theta=\left\{ \theta_i = b_s^{-4(i-1) / d}, i \in [1, 2, \ldots, d / 4] \right)$.
    \item Step 2: Calculate the rotation angles as the multiplication of the rotation frequency $\Theta$ and the token coordinates $\mathbf{x}=[\mathbf{x}^1,\mathbf{x}^2]\in\mathbb{R}^2$, then prepare $\boldsymbol{R}_{\Theta,m}^d$,
\begin{equation}
    \boldsymbol{R}_{\Theta,m}^d=\left(\begin{array}{ccccccc} \boldsymbol{R}_{\Theta,m}^{d/2,1}&\mathbf{0} \\ \mathbf{0}&\boldsymbol{R}_{\Theta,m}^{d/2,2}\end{array}\right)
\end{equation}
\begin{equation}
    \boldsymbol{R}_{\Theta, m}^{d/2,i}=\left(\begin{array}{ccccccc}
\cos \mathbf{x}_m^i \theta_1 & -\sin \mathbf{x}_m^i \theta_1 & 0 & 0 & \cdots & 0 & 0 \\
\sin \mathbf{x}_m^i \theta_1 & \cos \mathbf{x}_m^i \theta_1 & 0 & 0 & \cdots & 0 & 0 \\
0 & 0 & \cos \mathbf{x}_m^i \theta_2 & -\sin \mathbf{x}_m^i \theta_2 & \cdots & 0 & 0 \\
0 & 0 & \sin \mathbf{x}_m^i \theta_2 & \cos \mathbf{x}_m^i \theta_2 & \cdots & 0 & 0 \\
\vdots & \vdots & \vdots & \vdots & \ddots & \vdots & \vdots \\
0 & 0 & 0 & 0 & \cdots & \cos \mathbf{x}_m^i \theta_{d / 4} & -\sin \mathbf{x}_m^i \theta_{d / 4} \\
0 & 0 & 0 & 0 & \cdots & \sin \mathbf{x}_m^i \theta_{d / 4} & \cos \mathbf{x}_m^i \theta_{d / 4}
\end{array}\right)
\end{equation}
    \item Step 3: Update $q$ and $k$ in self-attention mechanism:
\begin{equation}
    f_{\{q, k\}}\left(\boldsymbol{h}_m, \mathbf{x}_m\right)=\boldsymbol{R}_{\Theta, m}^d \boldsymbol{W}_{\{q, k\}} \boldsymbol{h}_m
\end{equation}
, where $\boldsymbol{W}_{\{q, k\}}$ is a weight matrix to transfer token embeddings into queries and keys. 
\end{itemize}

\begin{figure}
    \centering
    \includegraphics[width=\linewidth]{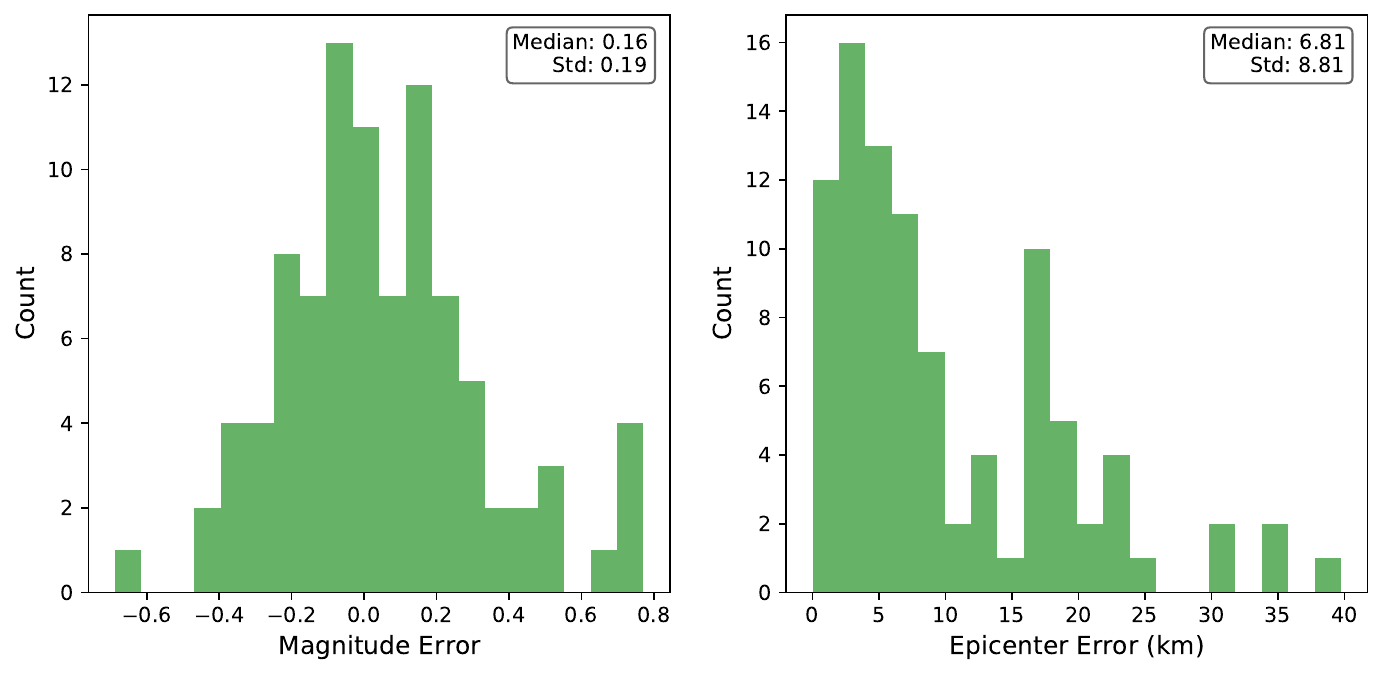}
    \caption{Histogram for magnitude and epicenter location error for 95 large earthquakes in 2023 ($M>$3.5), 4-s after P-wave arrival at the first triggered station.}
    \label{fig:EEW-hist-mag-loc}
\end{figure}

\begin{figure}
    \centering
    \includegraphics[width=\linewidth]{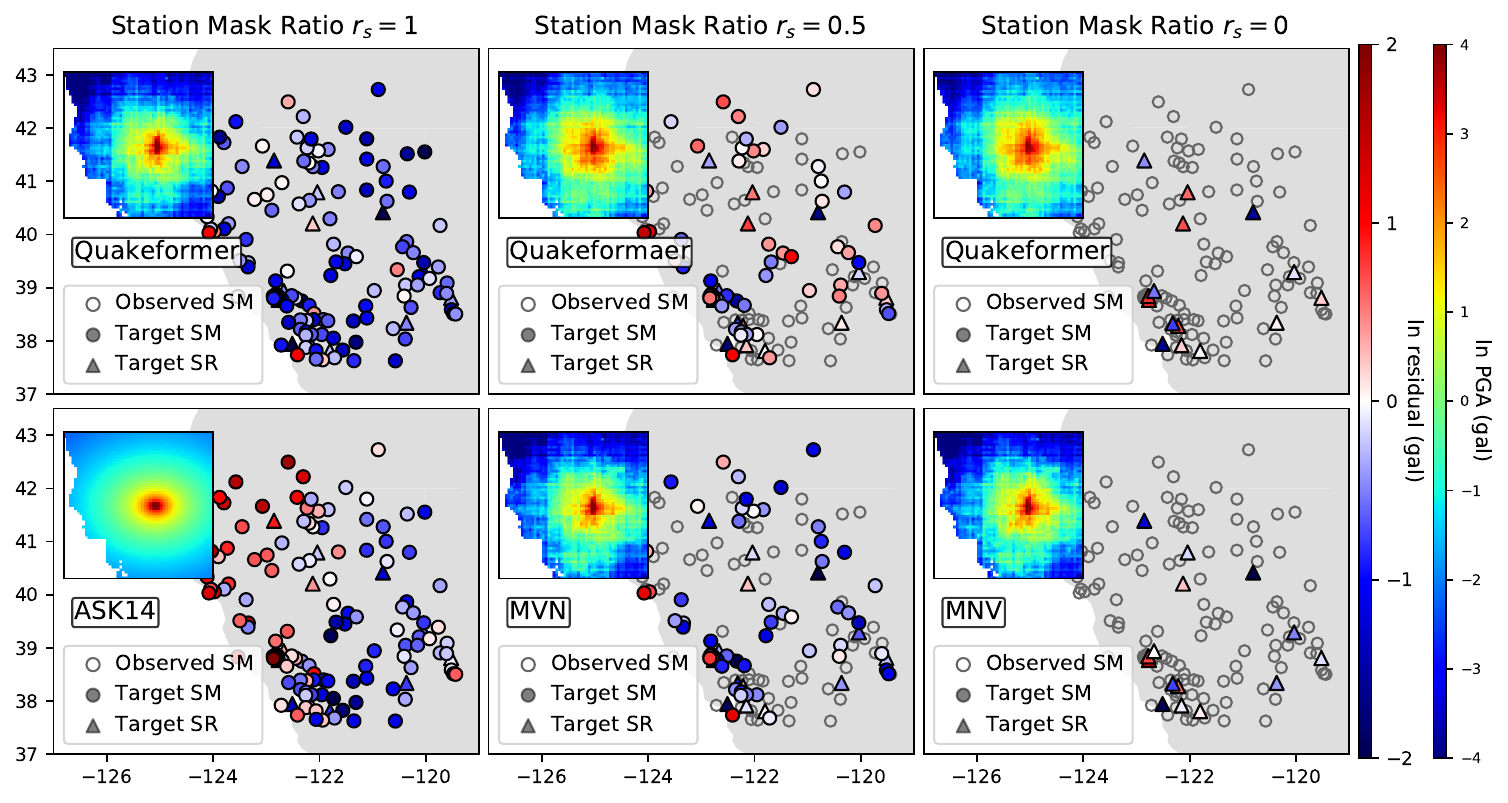}
    \caption{Forecasting-interpolation example result: nc73886731, \textbf{M}: 5.48 }
    \label{fig:interp-case-1}
\end{figure}

\begin{figure}
    \centering
    \includegraphics[width=\linewidth]{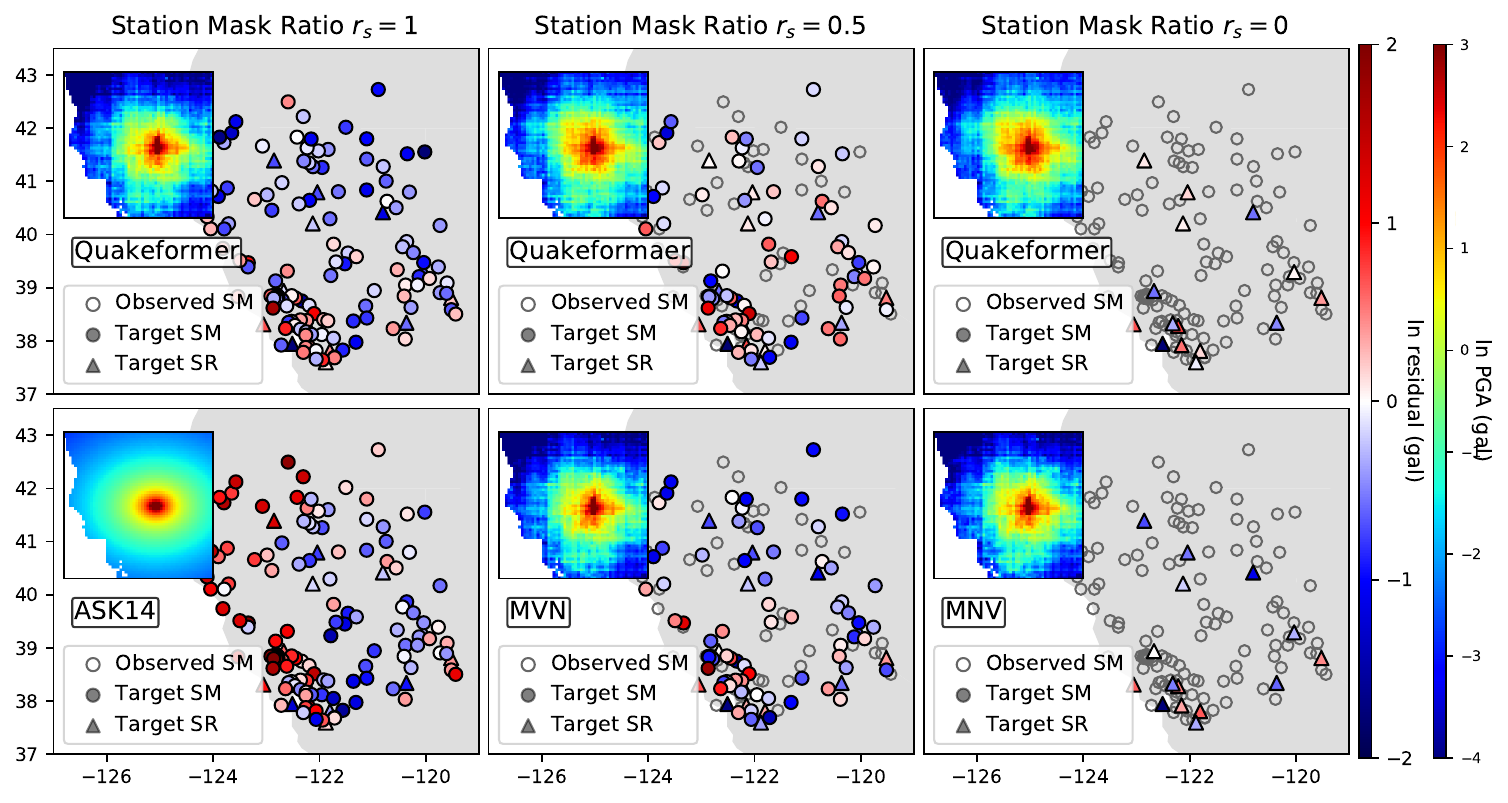}
    \caption{Forecasting-interpolation example result: nc73887046, \textbf{M}: 5.2 }
    \label{fig:interp-case-2}
\end{figure}

\begin{figure}
    \centering
    \includegraphics[width=\linewidth]{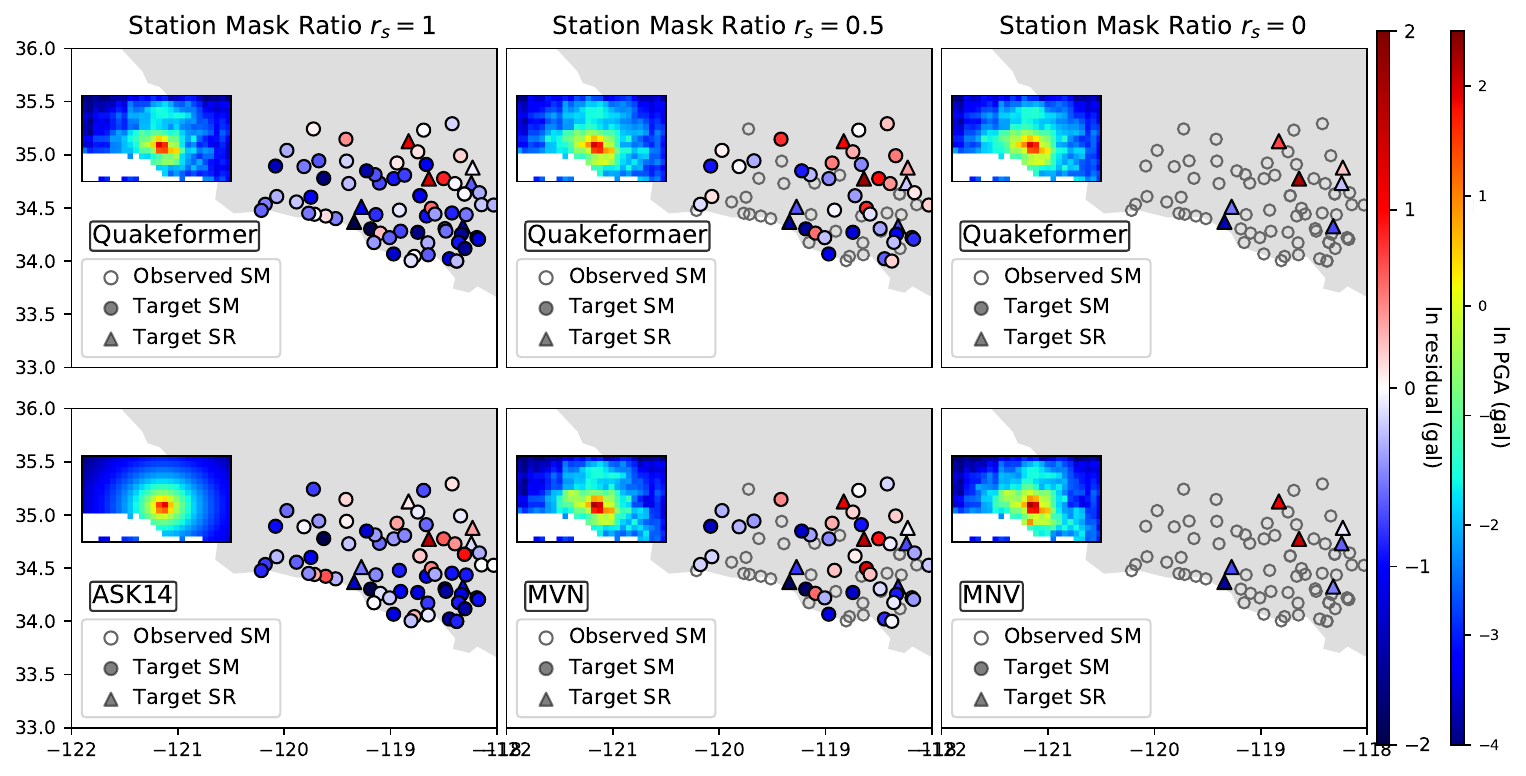}
    \caption{Forecasting-interpolation example result: ci40582824, \textbf{M}: 3.29}
    \label{fig:interp-case-3}
\end{figure}


\section{Baseline methods}
\subsection{Forecasting: ASK-14}
We use a simplified version of ASK14 in our experiment, similar with \citet{sahakian2018decomposing}, 
\begin{equation}
    \ln{PGA}=a_1+a_2Mag+a_3(8.5-M)^2+a_4\ln{R}+a_5\ln{\frac{V_{S30}}{V_{ref}}}
\end{equation}
, where $R=\sqrt{R_{rup}^2+c^2}$. Since the majority of our dataset is small-magnitude events, $R_{rup}$ is equivalent to the hypocentral distance here, in the unit of kilometer. $c$ is the finite-fault dimension factor, taken to be a constant 4.5\citep{Abrahamson2014ASK14}. $V_{ref}=760m/s$. The coefficients $a_1-a_6$ are determined by the least-squared(L2) inversion and single-mean models using the train events and train stations ($T_M+S_M$). Although this algorithm differs from the commonly used mixed effect(ME) model that constrains the between-/within-event residuals directly, the difference between them is within 0.02 natural log units in South California\citep{sahakian2018decomposing}. Therefore, the single model is still valid and more suited to test the extrapolation ability in $T_R+S_R$.
We apply the empirical equation to three regions, that is North California, South California, and the whole California. Table \ref{table_ASK14} list the model coefficients and R-squared for the three region. 

\begin{table}

\caption{ASK14 Model Coefficients and Regression R-squared}
\label{table_ASK14}
\centering
\begin{tabular}{cccccccccc}
\hline
Region   & $a_1$   & $a_2$   & $a_3$    & $a_4$    & $a_5$   & \begin{tabular}[c]{@{}c@{}}$R^2$\\ TM+SM\end{tabular} & \begin{tabular}[c]{@{}c@{}}$R^2$\\ TM+SR\end{tabular} & \begin{tabular}[c]{@{}c@{}}$R^2$\\ TR+SM\end{tabular} & \begin{tabular}[c]{@{}c@{}}$R^2$\\ TR+SR\end{tabular} \\ \hline
North CA & 3.7772 & 1.3672 & -0.0627 & -2.1926 & 0.1899 & 0.7600                                                               & 0.8090                                                               & 0.7808                                                               & 0.7607                                             \\
South CA & 1.0959 & 1.3290 & -0.0371 & -1.5507 & 0.0268 & 0.7170                                                               & 0.6812                                                               & 0.6702                                                               & 0.6579                                             \\
CA       & 2.6011 & 1.1556 & -0.0495 & -1.7799 & 0.0227 & 0.7100                                                               & 0.7348                                                               & 0.7055                                                               & 0.6734                                             \\ \hline
\end{tabular}
\end{table}

\subsection{Interpolation: MVN (Conditional Multivariate normal distribution Approach)}
In this study, we only discuss the distribution of PGA. MVN approach requires a ground motion forecasting model, a cross-correlation model, and the assumption that the residuals have an MVN distribution. We set random variables $\mathbf{Y}_1$ and $\mathbf{Y}_2$ as the prediction at unobserved locations and observed locations. First, ground motion forecasting models are used to provide the initial estimation. We use QuakeFormer as the forecasting model. The mean estimate provided by the forecasting for unobserved and observed locations are $\mathbf{\mu}_{\mathbf{Y}_1}$ and $\mathbf{\mu}_{\mathbf{Y}_2}$. The residual are treated as a linear mixed effects model: $\mathbf{Y}_i = \mu_{\mathbf{Y}_i}+E_i+W_i$. Then, the predicted PGA values at unobserved locations are conditioned upon the available nearby observations, resulting in the decrease of uncertainty \citep{Worden2018interpolationMVN}. The distribution of $\mathbf{Y}_1$, given by $\mathbf{Y}_2=y_2$ (in which $y_2$ is the observation treated as exact):
\begin{equation}
    \mu_{\mathbf{Y}_1 \mid \mathbf{y}_2}=\mu_{\mathbf{Y}_1}+\boldsymbol{\Sigma}_{\mathbf{Y}_1 \mathbf{Y}_{\mathbf{2}}} \boldsymbol{\Sigma}_{\mathbf{Y}_{\mathbf{2}} \mathbf{Y}_{\mathbf{2}}}^{-1}(y_2-\mu_{\mathbf{Y}_2})
\end{equation}
, where $\Sigma_{Y_i, Y_j}$ is the element of the covariance matrix, and given by:
\begin{equation}
    \Sigma_{Y_i, Y_j}=\rho_{Y_i, Y_j} \phi_{Y_i} \phi_{Y_j}+\tau_{Y_i} \tau_{Y_j}
\end{equation}
, where $\rho_{Y_i, Y_j}$ is the correlation, $\phi$ and $\tau$ are the within-event standard derivation and between- event derivation (details could be seen in Appendix \ref{Appendix residual}. Because in this study, we didn't estimate uncertainty by QuakeFormer, we assume $\phi$= and $\tau$=. As same with \citet{Worden2018interpolationMVN}, we employ the cross-correlation function defined by assuming the the spatial correlation is due to the distance between sites $i$ and $j$, 
\begin{equation}
    \rho_{\mathbf{Y}_i\mathbf{Y}_j}=\exp (-h/10)
\end{equation}
,in which $h$ is the distance between the two sites, in units of kilometers.

\subsection{Early Warning: Estimate Point Source (EPS)}
In ElarmS, the magnitude is estimated from the amplitude and frequency contents of the P-wave arrivals using empirical scaling relations. In this study, we adopt the scaling relationship between the displacement amplitude of vertical component $PGD$ and magnitude $M$ proposed by \citet{kuyuk2013global}. 
\begin{equation}
    M=c_1\log(PGD)+c_2\log(E)+c_3
\end{equation}
, where $E$ is the epicenter distance in kilometers and $PGD$ is in centimeters. We use the maximum value within 4 s of the P-wave, and stop the time window up to the S-wave arrival in our analysis. Similar with \citet{kuyuk2013global}, we filter the displacement waveforms by a 3 Hz Butterworth filter .We calibrate the scaling coefficients using a linear best fit to the California data sets and exclude records that 1) have epicenter distances greater than 100 km, 2) $PGA$<1e-5. The calibration result is, $c_1$=0.8663, $c_2$=1.4257, $c_3$=4.2853, with the $R^2$ 0.73. Since our data set is dominated by small events, the scaling relationship is not as significant as this in the larger event, which could also be found in \citep{kuyuk2013global}. In order to provide a single estimate for the event, magnitude estimates for all triggered station at each point in time are weighted based on the P-wave windows recorded so far. Note that, we use the catalog hypocenters in the analysis, although estimating the location of the epicenter is typically the first step in the ElarmS, which leads to inflated performance. 

\subsection{Early warning: The Propagation of Local  Undamped Motion (PLUM)}
The PLUM method predicts the ground motion intensity at a target location based on the intensity observed at seismic stations around the target location in radius $r$. The PGA estimation at time $t$ and site $k$ is $PGA_{t,k}^{\text {pred}}=\max _{i \in C_R}\left\{PGA_{t,i}^{\text {obs}}\right\}$. Due to the sparser station coverage in California, we used $r= 15$ km. The original version of PLUM also takes site amplification factors into consideration, but we didn't apply site effects in our analysis, since the previous findings shows that the site amplification terms have minor impact on the performance.

\section{Residual decomposition} \label{Appendix residual}
We use the same notation as \citet{baltay2017uncertainty}. Generally, the total residual $\delta_{ij}$ is calculated by $\ln{\mathbf{IM}_{ij}}=\mathbf{GMM}+\delta_{ij}$, with the total variability std.dev.$\left(\delta_{ij}\right)=\sigma$. This residual consists of a source-related term $\delta E_i$ expressing the difference between the specific source excitation and the average excitation in GMM, and a path/site term $\delta W_{ij}$ expressing the difference between the specific path/site effects and the average path/site effect in GMM. $\delta_{ij}=\delta E_i+\delta W_{ij}$, these two residuals are also referred to between-event residual and within-event residual. The between-event residual could be separated using the mean of all $\delta_{ij}$ belongs to event $i$, in which the number of station recordings from event $i$ is $N_i$, std.dev.$\left(\delta E_i\right)=\tau$:
\begin{equation}
    \delta E_i=\frac{1}{N_i} \sum_{j=1}^{N_i} \delta_{i j}
\end{equation}
Note that, because the residual $\delta$ at site SR is a complex combination of event effects as well as the model behavior of spatial prediction, we calculate the between event residual only use records belongs to SM. And for the test case when $r_s$ is not 1, a portion of seismic stations in SM will be randomly chosen as observation stations, and $\delta E_i$ is calculated using the residuals on the remaining SM sites. For each event in TR, we randomly generate the Station Mask with $r_s$ 10 times, and calculate $\delta E_i$ in this specific masking situation. The final $\delta E_i$ for these event is the mean of these tested $\delta E_i$.

PGA is dependent on earthquake stress drop and studies have identified regions in southern California with different stress drops. We hope to see if QuakeFormer could capture the region difference of epicenters, therefore, we separate a systematic, location-based event term $\delta L_l$ from $\delta E_i$ grouped by the earthquake clusters in Figure \ref{fig:event cluster in 2023}. $\delta L_l$ is defined as the average event term $\delta E_i$ in a particular region $l$:
\begin{equation}
    \delta L_l=\frac{1}{N_l} \sum_{i \in I}^{N_l} \delta E_i
\end{equation}

Then, the total between-event residual $\delta E_i=\delta L_l+\delta E_i^0$, in which $\delta E_i^0$ is the remaining random component. std.dev.$\left(\delta L_l\right)=\tau_L$ represents the uncertainty relates to souce locations; std.dev.$\left(\delta E_i^0\right)=\tau_0$ represents the aleatory effect.

The within-event residual $\delta W_{ij}$ is calculated as $\delta W_{ij}=\delta_{ij}-\delta E_i$, and could be further parsed into site term $\delta S_j$, path term $\delta P_{lj}$ and a remaining aleatory component $\delta W_{ij}^0$, $\delta W_{i j}=\delta S_j+\delta P_{l j}+\delta W_{i j}^0$. For simplicity, the sum of the last two term is also referred as single-station residual, $\delta WS_{ij}=\delta P_{l j}+\delta W_{i j}^0$. 
The site term is the average of $\delta W_{ij}$ at station $j$, in which $m_j$ is the number of event recorded at station $j$:
\begin{equation}
    \delta S_j=\frac{1}{m_j} \sum_{i=1}^{m_j}\left(\delta_{i j}-\delta E_i\right) .
\end{equation}

\begin{figure}
    \centering
    \includegraphics[width=\linewidth]{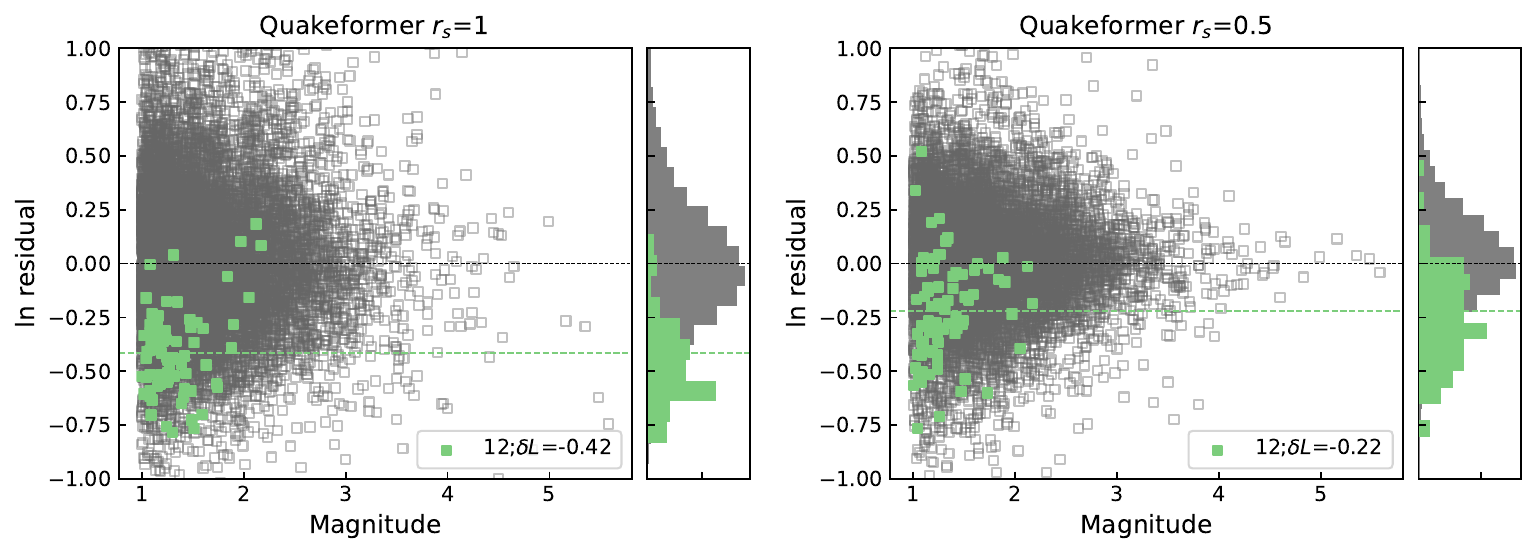}
    \caption{Between event residuals on  Cluster 12 for ground motion forecasting and interpolation tasks. a) is the residual distribution and histogram for QuakeFormer forecasting; b) is for QuakeFormer interpolation with station mask ratio $r_s=$ 0.5. The wave ratio for QuakeFormer is 0.}
    \label{fig:between event residual on cluster 12}
\end{figure}

\begin{figure}
    \centering
    \includegraphics[width=\linewidth]{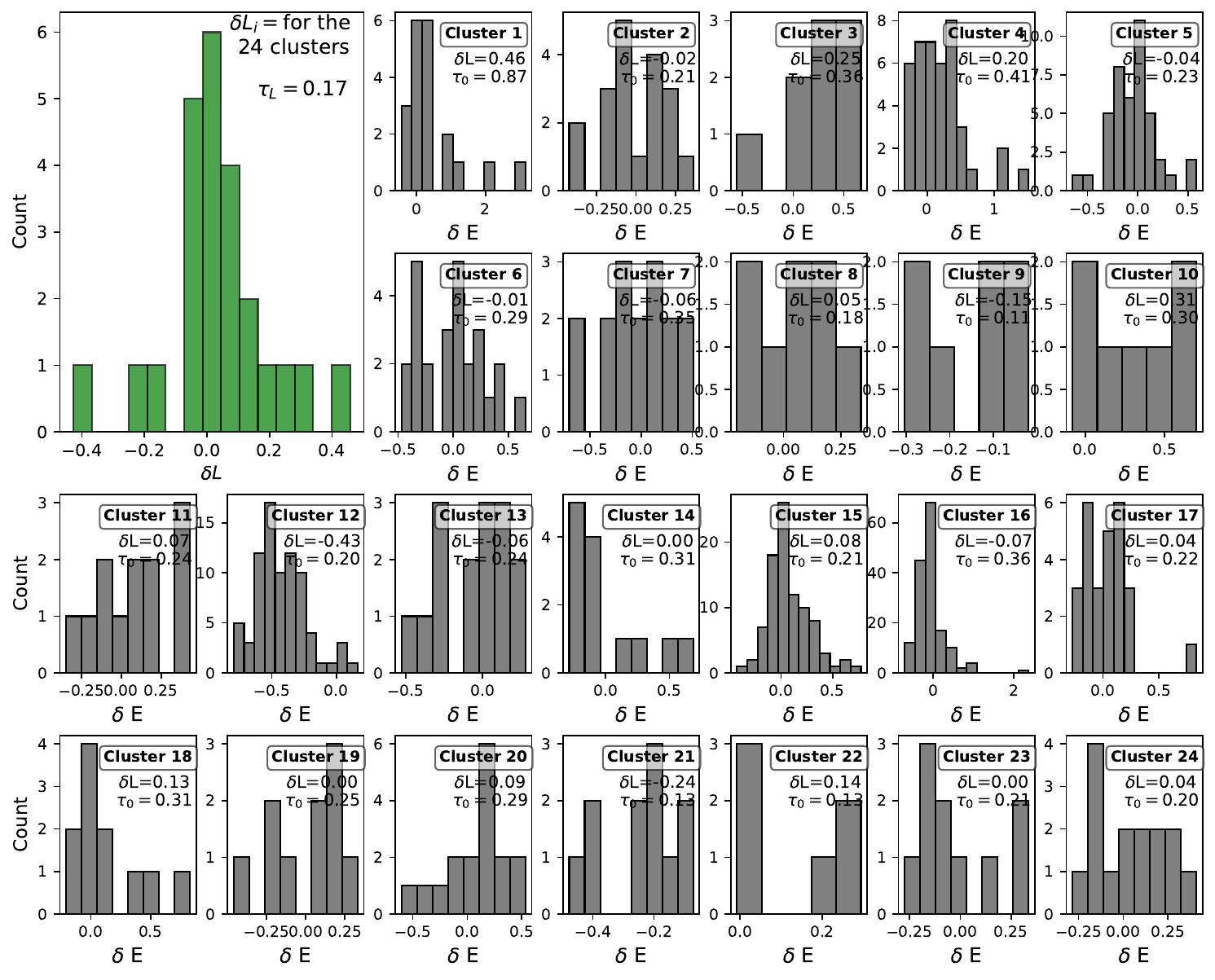}
    \caption{Histogram of $\delta L$ value for the 24 cluster. }
    \label{fig: event term histogram}
\end{figure}

\begin{figure}[h]
    \centering
     \includegraphics[width=0.75\linewidth]{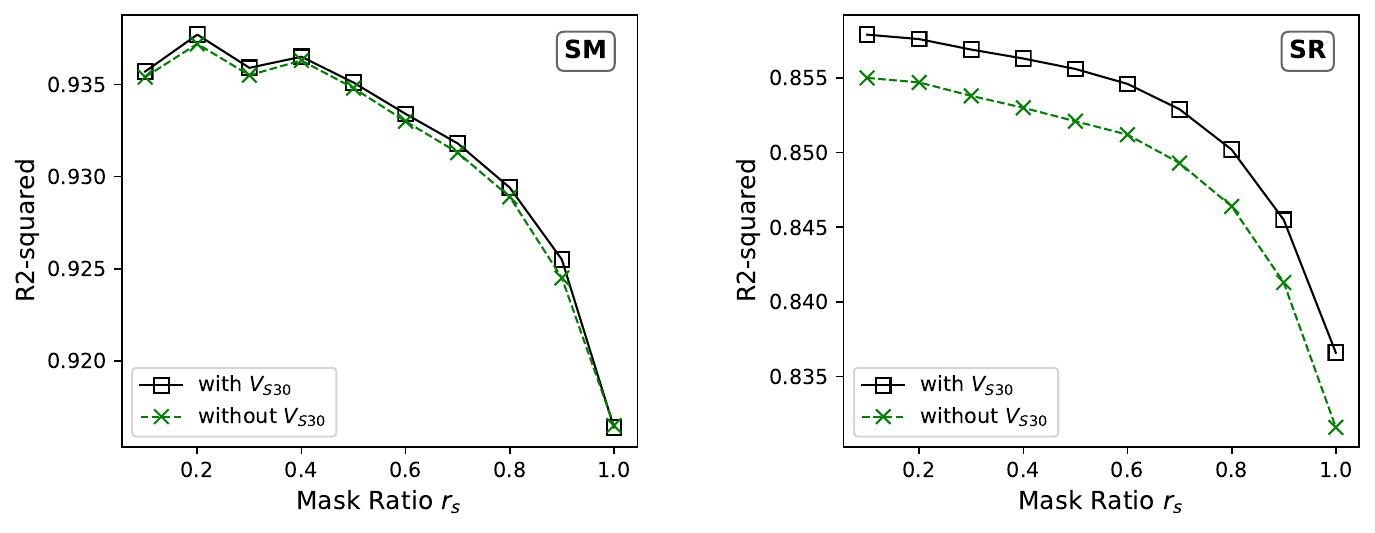}
    \caption{$R^2$ across different station masking ratio for QuakeFormer with (a) or without (b) $V_{S30}$.}
    \label{fig:R2 Vs30}
\end{figure} 

\begin{table}[]
\centering

\caption{Site term comparison with \citet{sahakian2018decomposing} (GMPE method). All station codes are from Caltech (CI) network, using high broad band high-gain seismometers (HH). All site terms are in ln units.}
\label{table: site term comparison}
\begin{tabular}{@{}ccc|ccc|ccc@{}}
\toprule
Site & QuakeFormer & GMPE  & Site & QuakeFormer & GMPE  & Site & QuakeFormer & GMPE  \\ \midrule
BAR  & -0.16       & 0.05  & LLS  & -0.02       & 0.24  & SDR  & 0.03        & 0.23  \\
BBR  & 0.04        & -0.45 & MGE` & -0.04       & 0.32  & SLB  & 0.19        & -0.78 \\
BBS  & -0.12       & -0.48 & MLS  & 0.07        & 0.31  & SLR  & -0.09       & 0.39  \\
BOM  & -0.01       & -0.14 & MSC  & 0.01        & -0.22 & SNO  & 0.24        & -1.3  \\
CRY  & 0.12        & -0.52 & MSJ  & -0.07       & 0.14  & STG  & 0.03        & -0.46 \\
CTC  & -0.05       & -0.2  & NSS2 & -0.12       & -0.33 & SVD  & 0.01        & -0.13 \\
DNR  & -0.09       & -0.07 & OLP  & 0.03        & 0.32  & SWS  & 0.02        & -0.78 \\
DPP  & 0.025       & 0.38  & PER  & -0.00       & -0.18 & TOR  & -0.01       & -0.66 \\
EML  & 0.01        & 1.15  & PLS  & -0.07       & -1.24 & WWC  & 0.00        & -0.64 \\
ERR  & 0.23        & 0.55  & RSS  & -0.05       & 0.66  & BOR$^{\dag}$  & -0.22       & 0.5   \\
FHO  & -0.02       & -0.37 & RVR  & -0.08       & 0.67  & DGR$^{\dag}$  & -0.15       & 0.39  \\
GOR  & 0.06        & -0.09 & SBPX & -0.01       & -0.51 &      &             &       \\
JEM  & 0.04        & 0.51  & SDD  & 0.06        & -1.31 &      &             &       \\ \bottomrule
\multicolumn{3}{@{}l}{$^{\dag}$ Sites belongs to SR}
\end{tabular}
\end{table}

\end{document}